\PassOptionsToPackage{table,xcdraw}{xcolor}

\documentclass[sigconf, natbib=true]{acmart}

\AtBeginDocument{%
  \providecommand\BibTeX{{%
    \normalfont B\kern-0.5em{\scshape i\kern-0.25em b}\kern-0.8em\TeX}}}

\usepackage{caption}
\usepackage{subcaption}
\usepackage{lscape}
\usepackage{soul}
\usepackage{booktabs}
\usepackage[table,xcdraw]{xcolor}

\copyrightyear{2023}
\acmYear{2023}
\setcopyright{rightsretained}
\acmConference[CHIIR'23]{ACM SIGIR Conference on Human Information Interaction and Retrieval}{March 19--23, 2023}{Austin, TX, USA}
\acmBooktitle{ACM SIGIR Conference on Human Information Interaction and Retrieval (CHIIR'23), March 19--23, 2023, Austin, TX, USA}
\acmDOI{10.1145/3576840.3578320}
\acmISBN{979-8-4007-0035-4/23/03}

\begin{document}

\title[The Evolution of Web Search User Interfaces - An Archaeological Analysis of Google SERP]{The Evolution of Web Search User Interfaces - An Archaeological Analysis of Google Search Engine Result Pages}


\author{Bruno Oliveira}
\affiliation{%
  \institution{Faculty of Engineering of the University of Porto}
  \country{Porto, Portugal}}
\email{up201605516@edu.fe.up.pt}

\author{Carla Teixeira Lopes}
\affiliation{%
  \institution{Faculty of Engineering of the University of Porto and INESC-TEC}
  \country{Porto, Portugal}}
\email{ctl@fe.up.pt}


\begin{abstract}
Web search engines have marked everyone's life by transforming how one searches and accesses information. Search engines give special attention to the user interface, especially search engine result pages (SERP). The well-known ``10 blue links'' list has evolved into richer interfaces, often personalized to the search query, the user, and other aspects. More than 20 years later, the literature has not adequately portrayed this development. We present a study on the evolution of SERP interfaces during the last two decades using Google Search as a case study. We used the most searched queries by year to extract a sample of SERP from the Internet Archive. Using this dataset, we analyzed how SERP evolved in content, layout, design (e.g., color scheme, text styling, graphics), navigation, and file size. We have also analyzed the user interface design patterns associated with SERP elements. We found that SERP are becoming more diverse in terms of elements, aggregating content from different verticals and including more features that provide direct answers. This systematic analysis portrays evolution trends in search engine user interfaces and, more generally, web design. We expect this work will trigger other, more specific studies that can take advantage of our dataset.
\end{abstract}




\maketitle

\section{Introduction}
The wealth of information available on the Web make search engines an essential tool nowadays~\cite{Amjad2014}. Web search engines have evolved a lot, and their user interface is no exception. Web search engines' front end is gaining importance in a scenario where relevance is becoming more difficult to evaluate by users, and their perception becomes more influenced by the user experience~\cite[p.~512]{Baeza-yates2011}.

Search interfaces support tasks from query formulation to selecting and understanding search results~\cite[p.~26-40]{Baeza-yates2011}. Typically, web search engines have a home page containing a search entry form in which the user types a query. Retrieval results are usually displayed as vertical lists on Search Engine Results Pages (SERP). For its richness, our study focuses on these pages. 
SERP have started as simple ``10 blue links'' pages. Although search engines have kept a consistent format of presenting search results, the information in SERP goes way beyond these links. The design of query interfaces and retrieval results display is an active area of research and experimentation. Although works provide an in-depth analysis of search user interfaces, such as the one from Hearst~\cite{10.5555/1631268}, the temporal evolution of SERP is understudied.  

Portraying the evolution of SERP contributes to preserving the history of web search user interfaces. Moreover, assuming these interfaces follow the general trends in web user interfaces, it contributes to the overall study of web interfaces. Broadly, it depicts trends in web design. This evolutionary analysis is also helpful for the interactive information retrieval community to understand better how the search interfaces have evolved in content, layout, and navigation and build upon this for further and deeper analysis. 

This study conducts an evolutionary analysis of Google's SERP during the last two decades of Google Search, analyzing the overall user interfaces over time and their elements. We chose Google for its popularity in the web search engine landscape. For this analysis, we have captured from the Internet Archive 5,653 Google SERP from 2000 to 2020.

This work makes several contributions. First, we present a systematization of the elements that appear or have appeared in a SERP, defining each and providing visual examples. This systematization can be helpful in future studies that have the SERP as their focus and allows the establishment of common terminology. Second, we analyze the evolution of the overall SERP and each element. Third, we propose a methodology that can be used to study different types of web search user interfaces (e.g., mobile ones) or user interfaces in other contexts. Fourth, this paper makes two resources available to the community: a dataset\footnote{Available at \url{https://doi.org/10.25747/991g-f765}} containing the screenshot and files associated with each extracted capture and a website\footnote{\url{https://bedgarone.github.io/serpevolution/}} summarizing the analysis. These resources can also be the input of further studies, inclusively done by researchers without advanced technological skills.


\section{Related work} \label{sec:relwork}

Research in search user interfaces focuses on their design and evaluation, either in broad or focused on query formulation, the presentation of search results, or even query reformulation. There are also works focused on personalization, information visualization, or domain-specific search interfaces such as mobile, social, or multimedia. There are whole books and monographs dedicated to this subject~\cite{10.5555/1631268, 10.5555/2502704, 10.5555/1824082, 6812556, Wilson2011, Bierig2021}. Despite such research, given our focus, we only describe works focused on analyzing the anatomy of SERP and, eventually, its evolution. This section does not cover works proposing or evaluating search interface components.

Höchstötter and Lewandowski~\cite{Hochstotter2009} address SERP composition and count the various elements' appearances. To the best of our knowledge, this was the first work to analyze the entire structure of the SERP. Besides, the authors examined the retrieved results, their sources, and types (e.g., organic results, advertisements, shortcuts). When the authors wrote this paper, advanced features in SERP were not widespread, which was not the case when Moran and Goray~\cite{Moran2019} studied the anatomy of SERP, defining the terminology for SERP elements. Nielsen Norman Group uses this terminology in several articles~\cite{ThreeKeySERP,Moran2019,FeiFei20,MoranAbandon}. 
In their `Search Patterns' book, Morville and Callender~\cite{10.5555/1824082}, apart from addressing the anatomy of the search process and related behavior, also list elements and principles of interaction design, illustrating many user interface design patterns around search websites. 


To the best of our knowledge, no works systematically collect and analyze SERP interfaces over time.

\section{Methodology} \label{sec:method}

The first stage of our work involves building a sample of SERP interfaces over time, a process described in Section~\ref{sec:whatcapture} and Section~\ref{sec:howcapture}. After collecting this sample, our attention focused on its analysis and automation, as described in Section~\ref{sec:automating}.

\subsection{What SERP have we captured?} \label{sec:whatcapture}

Google Search currently has 91.4\% of the market share~\cite{Chris2020}, a leadership that goes back to 2002~\cite{MktShare20102022,MktShare20002013}. In this context, we decided to focus our analysis on this search engine. 
This study will address desktop versions of Google Search from 2000 until 2020. A comparative analysis with the seconded ranked search engine, Microsoft Bing, is done in another work~\cite{serpevolutionShort} and is available on the study's website. 

The Internet Archive keeps snapshots and the respective HTML version of web pages over time. Its collection contains 588 billion web pages~\cite{InternetArchive1996}. 
Internet Archive provides 
the \emph{Wayback CDX Server API}, which allows complex querying, filtering, and analysis of captures. 
While filtering by URL, we can use a wildcard (*) at the end of the URL to specify the latter as a prefix and receive entries beyond the specified URL (e.g., \emph{\url{www.google.com/search?q=cookies*}}).

We found more than 195 thousand captures of Google SERP during two decades using the API. This large number of SERP and existing resource restrictions led us to devise a method to identify a smaller set of SERP.
To increase the likelihood of reaching pages with SERP element diversity, we have used a set of 129 most searched queries in the last 20 years, retrieved from Google Trends during the same period. This set\footnote{Available at \url{https://bedgarone.github.io/serpevolution/mostsearchedqueries}} contains the first search query from each available category, such as \emph{People}, \emph{Health}, or \emph{Electronics}. These queries include relevant terms often searched by users and trigger features in SERP. Hence, it is highly likely that SERP interfaces derived from these queries are richer and, thus, more relevant for this study than those generated by random searches. We decided to append these queries with the `*' wildcard while submitting them to the API to obtain more captures.

We noticed that some years had no captures using the most searched queries, which coincided with periods in which there were few captures from Google Search's domain. Hence, in those years, we collected all the available captures (\emph{all} method in Table~\ref{tab:dataset}). We also noticed that the last two years had much more captures (>10 thousand). Therefore, in 2019 and 2020, we restricted the URL submitted to the API to those containing queries shorter than 37 characters. Considering that an English word has, on average, about five characters~\cite{avgwords, annWords}, the 37 characters are the equivalent of 6 words plus spaces between them, which is more than the two to three words that a query typically has~\cite{Belkin2003, Jansen2005, Jansen2007}. This restriction excludes more specific queries that are probably less useful to the plurality of interfaces. Our sample has 5.653 captures. The last column of Table~\ref{tab:dataset} has ordered lists with the number of captures per year.


\begin{table}[h]
\small
  \caption{Method used to collect the captures, maximum length of the query (search URL), the width of the screenshot, and the number of captures extracted per year.}
  \label{tab:dataset}
  \begin{tabular}{ l c c c l}
    \toprule 
      & method & max. length & width & \#captures per year\\
    \midrule
    \textbf{2000 - 2002}& queries & - & 800px & 200, 3, 23\\
    \midrule
    \textbf{2003 }& queries & - & 1024px & 231\\
    \midrule
    \textbf{2004 - 2008}& all & - & 1024px & 12, 0, 200, 0, 26\\
    \midrule
    \textbf{2009 }& queries & - & 1024px & 11\\
    \midrule
    \textbf{2010 }& all & - & 1024px & 78\\
    \midrule
    \textbf{2011 }& queries & - & 1024px & 7\\
    \midrule
    \textbf{2012 - 2018}& queries & - & 1366px & 57, 975, 30,\\ 
    & & & &89, 172, 192, 548\\
    \midrule
    \textbf{2019 }& queries & 37 char & 1366px & 171\\
    \midrule
    \textbf{2020 }& queries & 37 char & 1920px & 2628\\
    \toprule
   \end{tabular}
\end{table}

\subsection{How have we captured SERP?} \label{sec:howcapture}
\begin{figure}
    \centering
    \includegraphics[width=1.0\columnwidth]{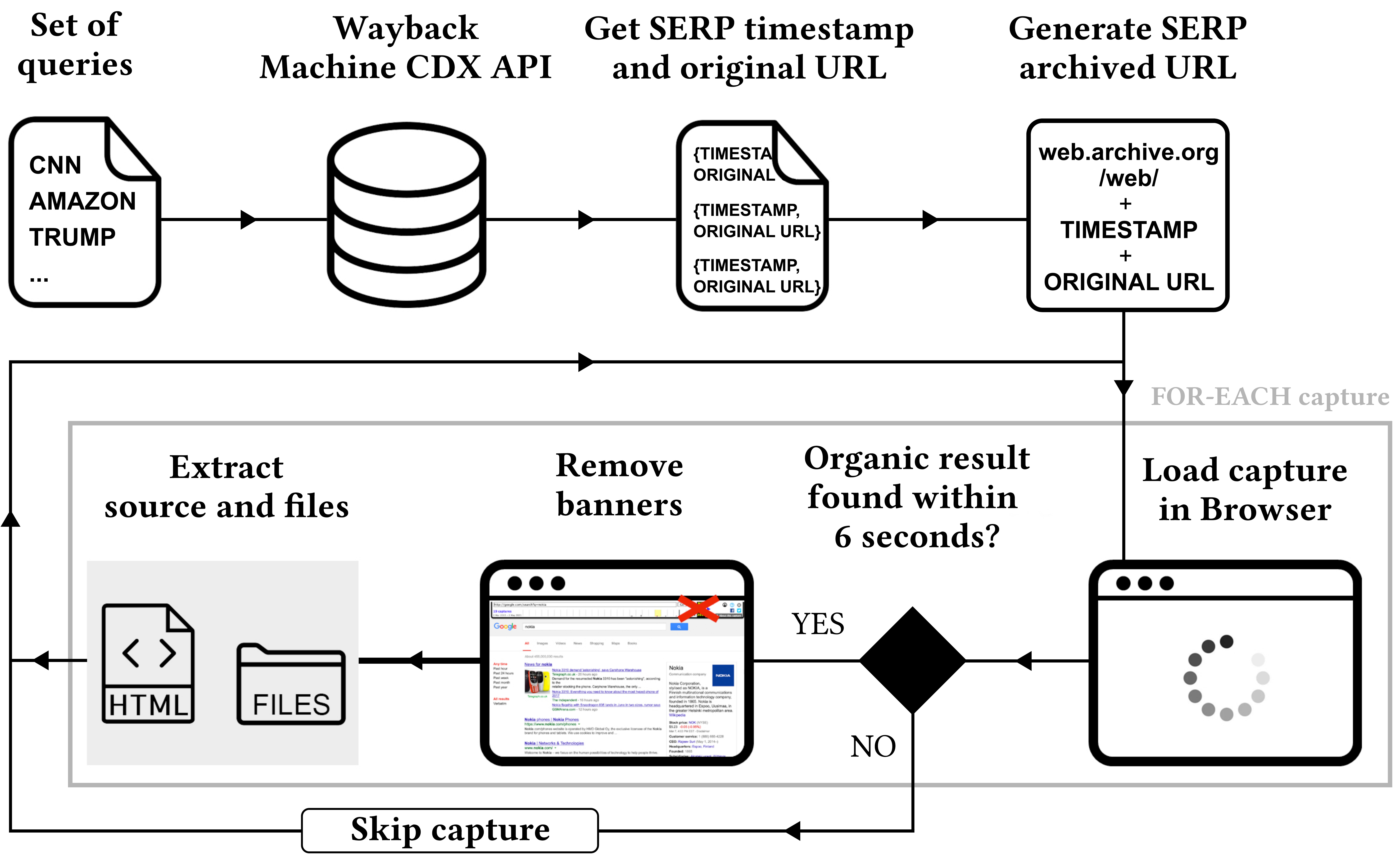}
    \caption{Extracting captures procedure}
    \label{fig:diagram}
\end{figure}

We used Python and Selenium Webdriver to visit each capture online, check if the capture was valid, save the HTML version, and generate a screenshot. The capture process is shown in Figure~\ref{fig:diagram}. The \emph{original URL} is the URL of the original SERP (e.g., \url{google.com/search?q=photography}), while the \emph{archived URL} is the URL of its archived version (e.g., \url{web.archive.org/web/20160125203434/www.google.com/search?q=photography}).

Some captures with an HTTP OK status code were not considered valid. Some are inexistant, showing a contradictory message of \emph{URL not captured}, while others are defective (e.g., incomplete interfaces without search results). To automatically check the validity of each capture, we try to find a result entry, the element that cannot lack in a SERP. Captures from SERP tabs other than the general first page, identified with ``tbm'' in the URL, were discarded for being outside this work's scope. We raise a timeout exception after 6 seconds, the time we empirically considered sufficient to load the capture in the browser. In these situations, the program will skip the capture. Before downloading the page, we still remove graphical elements from Internet Archive, such as its information and donation bars. Some of the captures present other distracting banners and overlapping parts of the interface, such as the ones related to cookie consent. We removed all the identified ones and extracted the source and associated files.

The process concludes with generating full-height screenshots of every HTML version opened in another browser instance in headless mode. We produced screenshots considering the most popular screen size at the time of the capture, as stated by the statistics~\cite{TeoalidaScreen}. We only considered the width, shown in Table~\ref{tab:dataset}, because SERP height is highly variable. The dataset with all the extracted captures is available online\footnote{\url{https://doi.org/10.25747/991g-f765}}. 

\subsection{How have we analyzed SERP?} \label{sec:automating}

\begin{figure}[h]
    \centering
    \includegraphics[width=1.0\columnwidth]{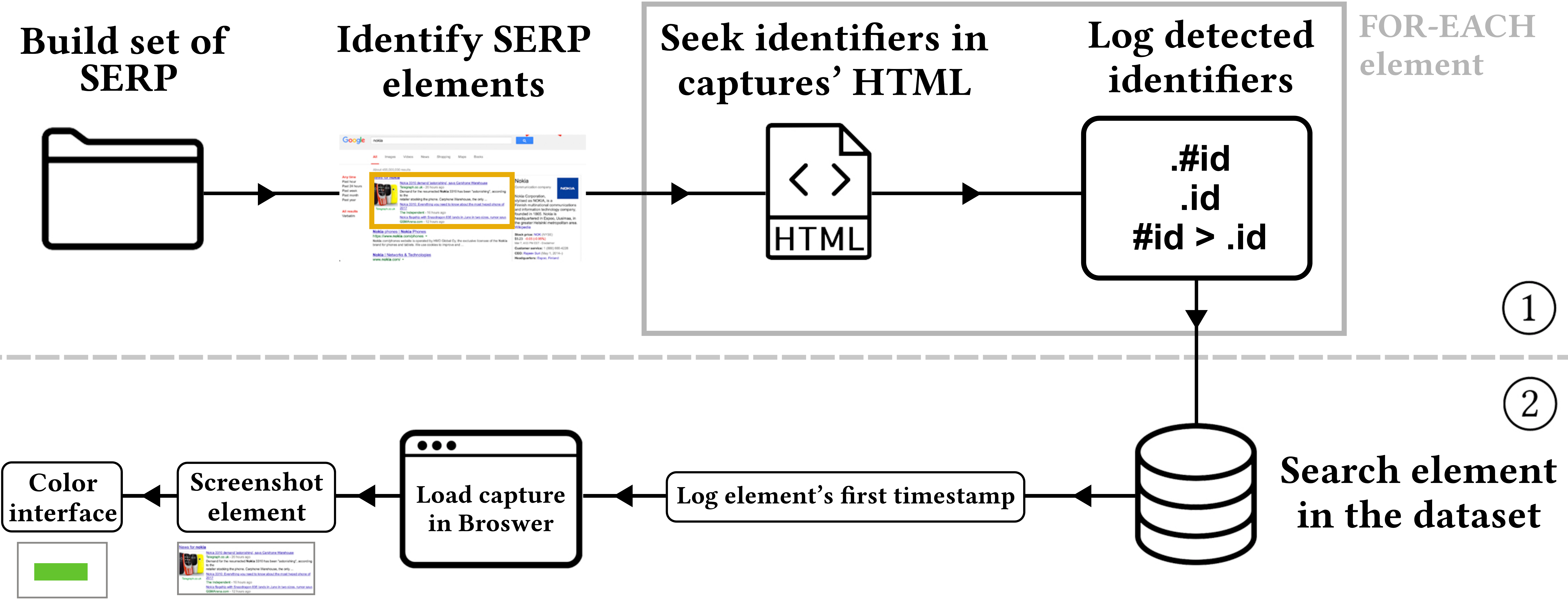}
    \caption{Detection and analysis of elements procedure}
    \label{fig:diagramanaly}
\end{figure}

The analysis process included two main stages, as shown in Figure~\ref{fig:diagramanaly}. First, we have extracted a sample of captures from the primary dataset to identify SERP elements. For each month with captures in the primary dataset, we manually looked at the screenshots of that month's captures and selected the capture with the most features. In the end, this set included 117 captures, with which we visually identified SERP elements. We manually analyzed each element's source code, looking for identifiers to locate the element in a later automated process. Element identifiers consist of HTML classes, ids, tags, or a combination of these using CSS selectors (e.g., \emph{`.knowledge-panel'}, \emph{`\#tads'} or \emph{`\#newsbox'}).
All the encountered identifiers were logged and listed on the website\footnote{\url{https://bedgarone.github.io/serpevolution/elements}}.

In the second stage, we automated the detection of these elements over time, allowing the exploration of a more significant number of cases. Finding an element with these identifiers triggers a function that stores the date of the element's appearance in a log file. We imposed no limit of captures per month to register the element's appearance, as the computation permits a full dataset scan in an acceptable time. The function also receives the element's upper-left corner coordinates, width, and height, generating and saving its image in the element's folder. Contrary to the element's timestamp, we imposed a limit of 15 captures per month while screenshotting to reduce and make the scan time feasible. We estimate that 15 monthly samples are enough to capture the possible changes of an element. 

Following a similar procedure, we automatically used the identifiers to detect and color the web page's targeted areas. We used Python, Selenium Webdriver, and BeautifulSoup to scrape every HTML capture to identify and generate transparency-colored images for each category of elements. We generated these images in a headless browser with a 1920px width, regardless of the capture's width. This simplification does not affect the final result because the elements in the interfaces do not move dynamically as the width increases or decreases. We have not constrained the height in this generation process. Due to the size of the dataset, we imposed a limit of 15 elements per month.

\begin{figure}[h]
    \centering
    \includegraphics[width=1.0\columnwidth]{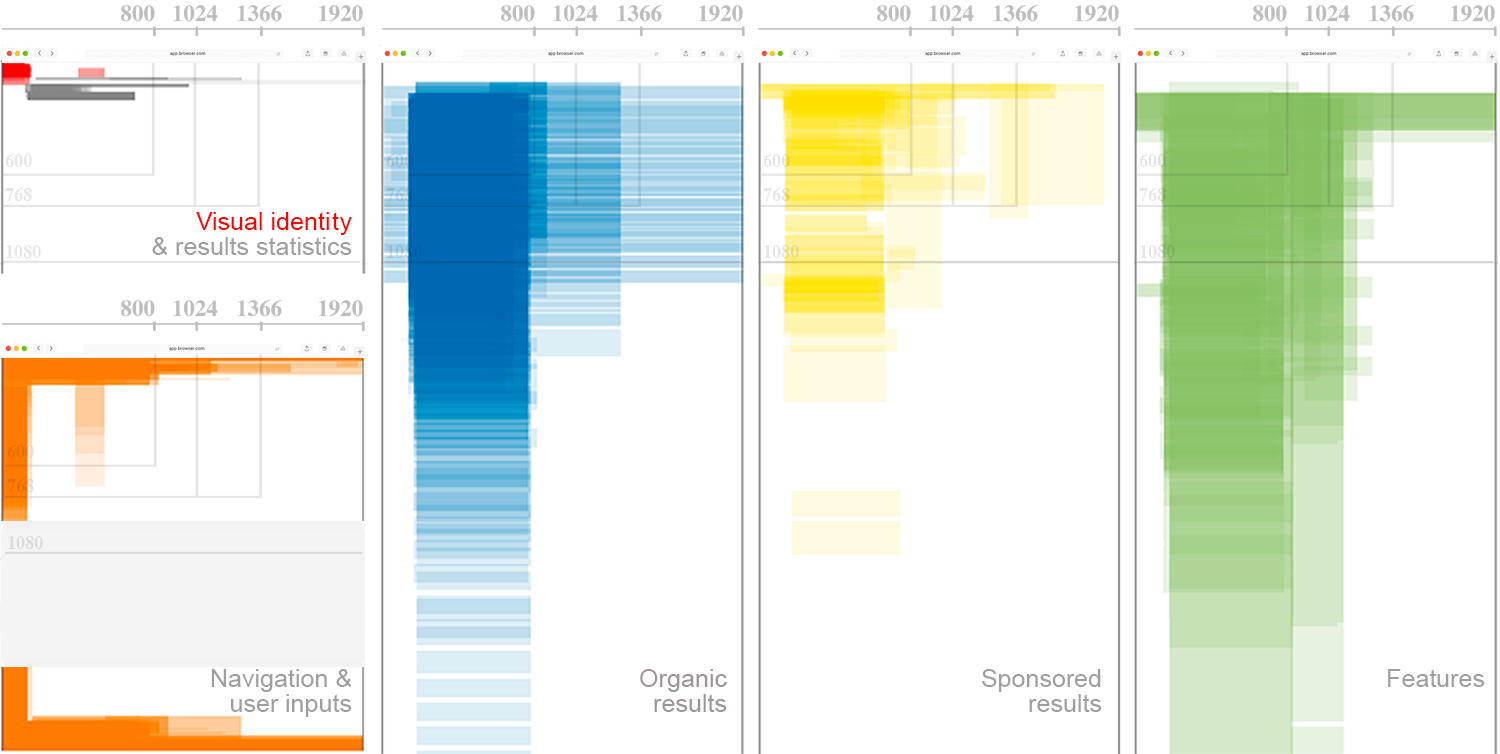}
    \caption{Transparency-colored overlaying results for each category}
    \label{fig:patterns_all}
\end{figure}

We overlayed all the individual images from single captures for each category of elements, which allows the overlay to enhance the most common areas while leaving the others almost unnoticeable. The overlaying process uses the upper-left corner as the reference for image alignment. \emph{Navigation \& user inputs} includes elements in and next to the footer, where common areas were not evident due to the page height's variability. In this case, to generate and correctly overlap the footers of the interface, we considered a height value of 600px for the footer, cropping it from the bottoms of the interface. The 600px height was estimated after a visual analysis of SERP pages and included a margin of error to ensure we covered all elements under study.
Thus, the orange result displayed in Figure~\ref{fig:patterns_all} is trimmed in the middle and combines those two capturing steps. We will analyze the positioning of every category of elements in the next section. An animated version of each result in this Figure is available on the website\footnote{\url{https://bedgarone.github.io/serpevolution/layout}}, permitting us to observe how the positioning of the elements' categories in SERP changed over time. 

\section{Evolution of SERP Elements} \label{sec:componentsevol}

We present each element's description and analyze its period of presence and positioning in SERP (also displayed on the website\footnote{\url{https://bedgarone.github.io/serpevolution/elements}}). Moreover, we analyze each element's evolution regarding content, graphics, navigation, and their relation with user interface \emph{design patterns}. These patterns are problem-oriented and generally repeatable solutions to usability problems in interface and interaction design~\cite{eelke,Tidwell2020}. 

\subsection{Visual identity \& search statistics}
Considering Google's visual identity, each logo version has kept its position and size with rare variations. Figure~\ref{fig:patterns_all} shows the overall position of the logo in red. Some interfaces are exceptions, offering a more significant left margin and a right-shifted logo and search statistics. Figure~\ref{fig:patterns_all}, in gray, shows how search statistics appear consistently below the search query or navigation bar, either left-aligned, right-aligned, or justified. Statistics included the number of results seen per page in the first decade. Later, Google removed this information and kept only the estimated number of results. Details about the logo evolution and the content about search statistics can be seen online\footnote{\url{https://bedgarone.github.io/serpevolution/design}}.

\subsection{Navigation \& User Inputs}\label{sec:navinputs}
\begin{figure}[h]
    \centering
    \includegraphics[width=1.0\columnwidth]{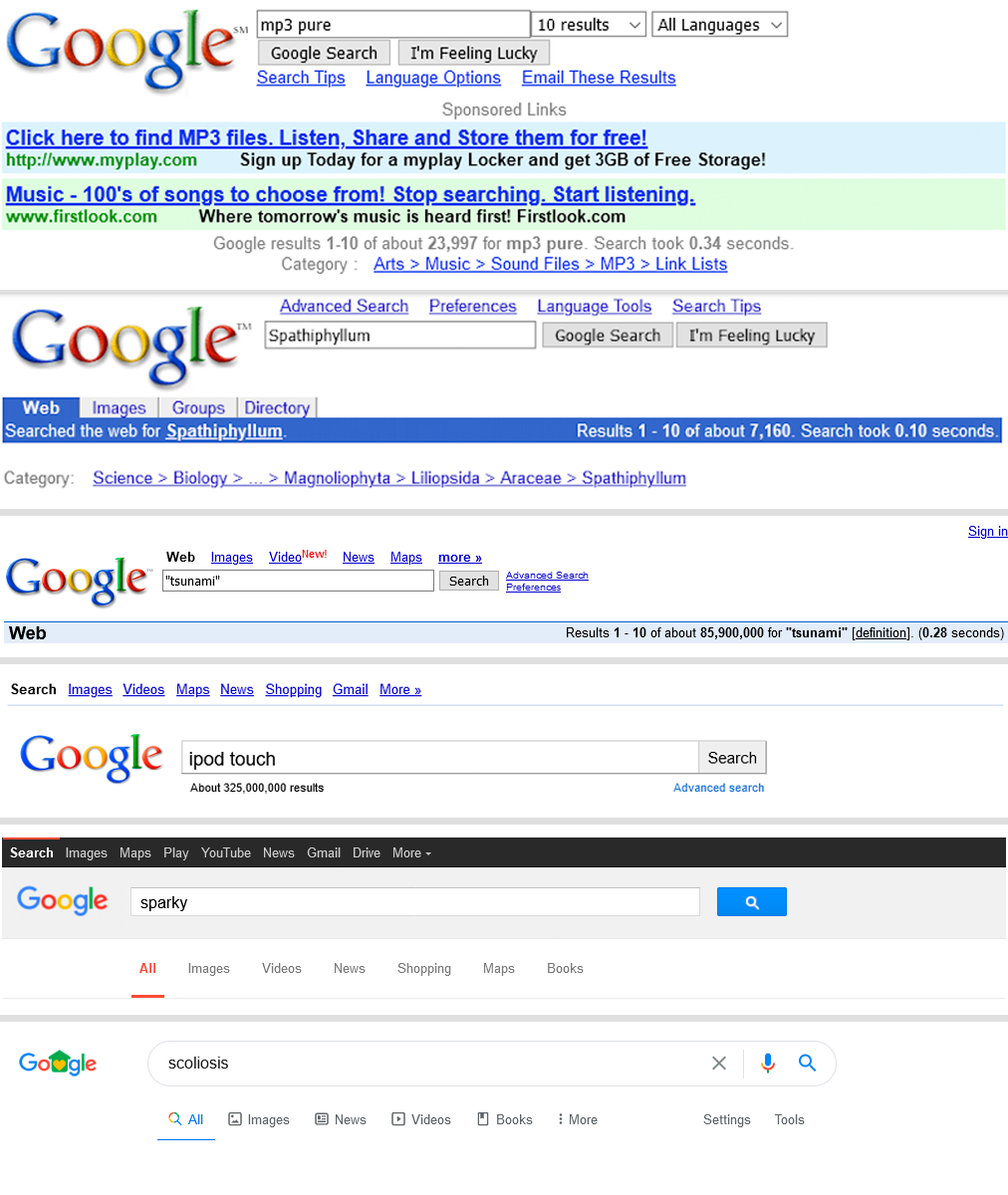}
    \caption{SERP headers neatly from 2000, 2001, 2006, 2012, 2017 and 2020}
    \label{fig:allnav}
\end{figure}

\begin{figure}[h]
    \centering
    \includegraphics[width=1.0\columnwidth]{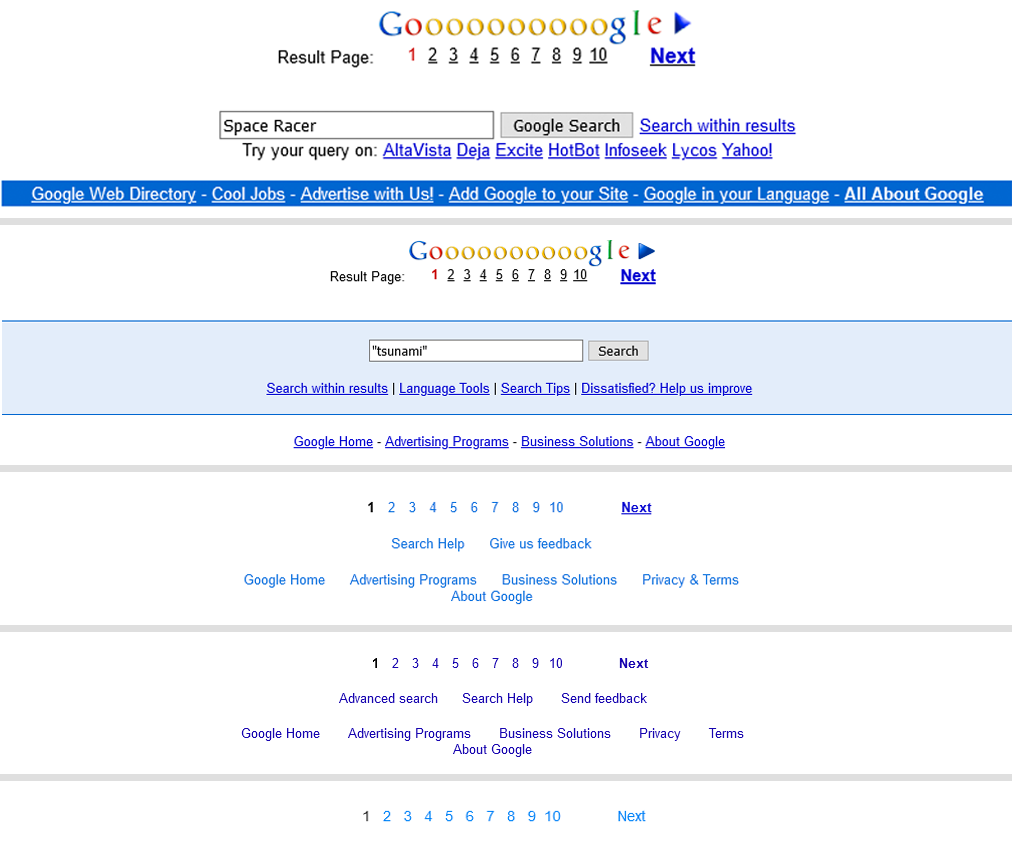}
    \caption{SERP footers neatly from 2000, 2006, 2012, 2017 and 2020}
    \label{fig:allfoot}
\end{figure}

Figures~\ref{fig:allnav} and \ref{fig:allfoot} display the main stages of how the search box and its surroundings evolved. The left-aligned query bar has also marked its place at the top of the page. Yet, in the early phases, it also appeared at the bottom, as seen in the 2000 and 2006 screenshots of Figure~\ref{fig:allfoot}. The \emph{Input Prompt} design pattern has always been applied to User Inputs.

We notice a change in the width of the entry form after 2006, which may suggest an encouragement to the formulation of longer queries~\cite{Karlgren2000, Belkin2003}. This change is in line with experimental evaluations where query length is positively related to effectiveness in the IR task~\cite{Belkin2003, Kurland2021}. Since query formulation influences effectiveness more than algorithmic factors~\cite{Kurland2021}, it makes sense to encourage users to do so.
We can also notice the appearance of a way to specify the query via spoken commands in the latest screenshot. This appearance aligns with the increasing attention conversational search interfaces have received~\cite{anand_et_al:DR:2020:11983}. Although works show promising results in incorporating visual elements in query formulation~\cite{Svarre2022}, there is no sign of them in Google Search.

The search box and the current query are always visible for the searcher, as recommended~\cite{Wilson2011}, since 2019. This visibility occurs even when the user scrolls down and the top of the SERP is no longer visible. Before 2019, the query was always available in the search box, but this box disappeared if the user scrolled down. Because we cannot interact with past SERP versions, we cannot analyze interaction features such as auto-complete or auto-correct.

In 2000, as seen in Figure~\ref{fig:allnav}, the main buttons for this category were `Google Search' and `I'm feeling lucky'. It was possible to select, directly in a dropdown near the search box, how many results should be shown, the language intended for the results, and an option to send the retrieved results by email. In the screenshots of 2000 and 2001 presented in Figure~\ref{fig:allnav}, the \emph{Category Hierarchy} feature is also visible showing one or more categories related to the query, probably obtained from Google Web Directory. This feature, available until 2004, located these categories in more general areas using \emph{breadcrumbs}, a design pattern that linearly specifies hierarchy levels leading to the current subject or page~\cite{Tidwell2020,toxboe}. At the bottom of the interface, there was also an option to search within the results and links to trigger the query on other Web search engines. Other lesser relevant links would point to SERP experience (e.g., language, search tips).

Although nonexistent for several years, it is possible to notice in Figure~\ref{fig:patterns_all}, in orange, the significant presence of the left navigation bar during some years of SERP history. This left column in orange is naturally interrupted in the middle area of the interface, trimmed as explained in Section~\ref{sec:automating}.

In 2001, Google introduced the tabs bar in the shape of Module Tabs used when content is groupable and there is no room for everything. Modules of content are divided into small tabbed areas with only one visible at a time, allowing the user to click on tabs to reveal other modules~\cite{Tidwell2020,toxboe}. Tabs don't need to be literal tabs and don't have to be at the top of the stack of modules~\cite{Tidwell2020}. The modules were Web, Images, Groups, and Directory. In 2002, Google added the News tab. In 2006, the tabs were replaced by simple links above the search query. This year, the \emph{Video}, \emph{Maps}, \emph{Froogle}, and \emph{More} tabs were introduced. In 2008, the tabs bar went to the very top of the page, and the Shopping and Gmail tabs were included. In 2010, the left sidebar was introduced, complementing the interface with other tabs and information such as location and results filtering. In 2012, Google removed the bottom query bar. In 2013, the tabs bar was displaced underneath the main query bar, maintaining some links and changing others. At the same time, in the left sidebar, only the results' filtering options were available before Google removed this sidebar in 2014. Finally, in 2015, the tabs bar below the query bar became the only existing one.

There is also a usual space for sign-in and user account information on the right side of the screen. At the bottom of the page, two areas are noticeable: pagination, aligned to the center of the results container, and the footer, at the very end, covering all the width. These areas are visible in Figure~\ref{fig:patterns_all}. The \emph{Pagination} design pattern has been applied to Navigation since the beginning of SERP. Pagination breaks up the long results list into different SERP, loading them one at a time~\cite{Tidwell2020}.

\subsection{Organic Results}\label{sec:orgres}
Organic results are retrieved based on the document's content and the overall retrieval algorithm. Regarding their positioning, in Figure~\ref{fig:patterns_all} (blue), there is a more substantial presence of colored frames in the area where results are typically included (henceforth called results container), with a greater focus on the visible area. By visible area, we mean the interface area that can be seen without scrolling. Information that needs scrolling to be accessed is in a scrolling area. Over time, it is noticeable that SERP pages have increased their height due to the vertical decrease of color intensity, revealing results in lower page regions. Part of the results is slightly shifted to the right, referring to interfaces with a left-side navigation bar. It is possible to observe two other very tenuous sets. One with more centered results since the interface from 2010 to 2012 adjusted elements' position according to the width and central axis of the screen rather than a left alignment. Some frames cover the entire interface width, not because the content was that large, but because, in the early days, HTML divisions (\emph{div}) typically spanned across the complete width of the viewport because Document Object Model (DOM) trees were less deep. The viewport consists of the visible area of an interface on a display device.

\begin{figure}[h]
    \centering
    \includegraphics[width=1.0\columnwidth]{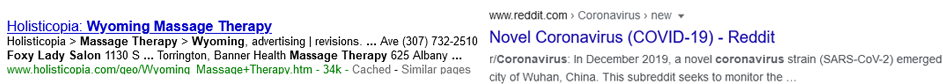}
    \caption{Regular result from 2003 (left) and 2020 (right)}
    \label{fig:regularres}
\end{figure}

Organic results can be regular or enriched. The structure of \textbf{regular results}, seen in Figure~\ref{fig:regularres}, started as a basic block of the page title, snippet, and URL links for \emph{similar} or/and \emph{cached} pages for the result. In 2013, Google hid these links in a dropdown, only visible by its arrow icon until now. In 2018, Google introduced a link to translate the result. 

\begin{figure}[h]
    \centering
    \includegraphics[width=1.0\columnwidth]{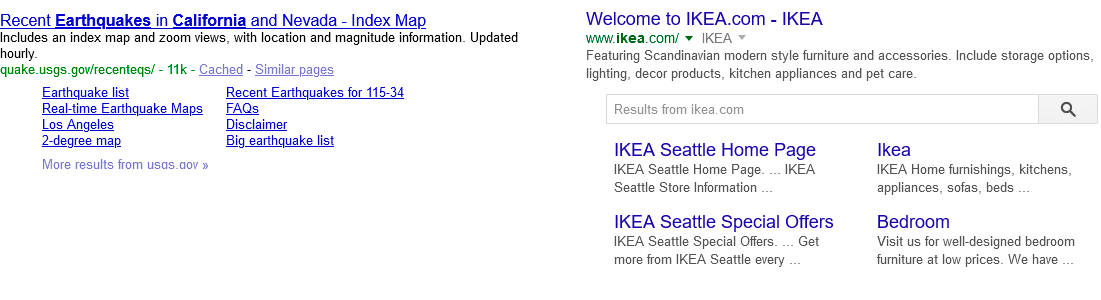}
    \caption{Enriched result from 2009 (left) and 2016 (right)}
    \label{fig:enrichedres}
\end{figure}

The \textbf{enriched results}, seen in Figure~\ref{fig:enrichedres}, are variations of regular results, with extra elements below the title, snippet, and URL, giving some additional information to the user. This element can have greater visibility and, in turn, a higher click rate~\cite{Rosu2020,Chakrabarti2009,Haas2011}. It appeared for the first time in 2008, lasting until now. Its positioning is consistently at the top of the results container. Initially, the extra content included two columns of site links pointing to sub-pages of the result's domain. These links, named \emph{quicklinks}, are navigational aid that attempts to take the users to the content they want quickly~\cite{Chakrabarti2009}. In 2009, it started highlighting structured data, such as reviews and ratings for products and services, based on experiments that showed that users find value in this new data~\cite{RichSnippets2009}. In 2010, Google enhanced each site link with a short description. In 2016, a search bar was introduced so that the user could search the result's website directly from SERP.

In both regular and enriched results, query term highlighting with boldfacing has been applied since 2006 to improve the usability of search results listings~\cite{Marchionini1995, Lesk1997, Clarke2007}. The dominant colors are blue for the title, green for the URL, and black/grey for the snippet. About 2011, the URL moved above the snippet. Google underlined the title until 2014. In 2020, the URL changed its color to gray and moved above the title. The new position of the URL may have been influenced by the importance of URL and domain names in evaluating the credibility of a result~\cite{Lin2011, Schwarz2011}. In 2019, the URL started being displayed as a breadcrumb, which may have been motivated by research that concluded that long and complex URL negatively impact clickthroughs~\cite{Clarke2007}.

In both types of organic results, the snippet length has not visibly changed, with most of the snippets having two lines, the suggested size for informing the user, and including as many results as possible in the visible area~\cite{Wilson2011}. As most of the queries used for data collection are associated with informational tasks, we cannot conclude if Google adjusts the length of the snippet to the type of query (e.g., informational, navigational) as suggested by existing research~\cite{Cutrell2007}. 

\subsection{Sponsored Results}
\begin{figure}[h]
    \centering
    \includegraphics[width=1.0\columnwidth]{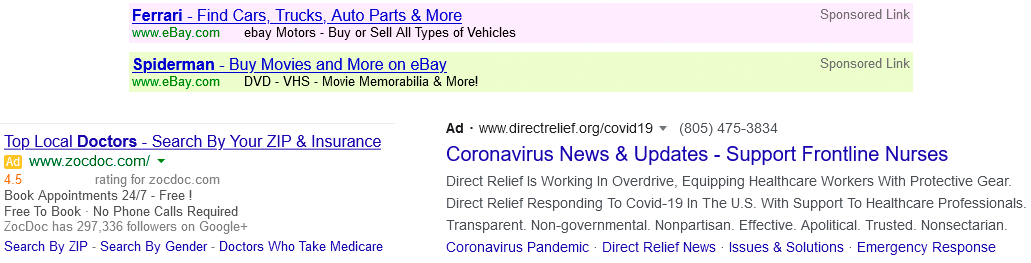}
    \caption{Textual ads from 2002 (top), 2014 (left) and 2020 (right)}
    \label{fig:textualadsres}
\end{figure}

As seen in Figure~\ref{fig:textualadsres}, \textbf{textual ads} are short advertisements that appear alongside organic results. These entries are focused on commercial intent, combining relevance with revenue, and are usually manually crafted, a standard in the advertisement industry~\cite{Danescu-Niculescu-Mizil2010}. This element has been present in SERP since the first day, lasting until now. This element's content and evolution are identical to the ones of \emph{Regular Results}, except these were initially marked with a specific tag, `Sponsored link'. In 2012, some of the results started to include \emph{quicklinks}, although less noticeable than in \emph{Enriched results}. Until 2014, a different background color distinguished these elements from organic results. Scarce occurrences in 2017 also had a colored background. In 2014, the tag `Ad' substituted the previous one, having a yellow background color, while from 2016 onwards, it was green. In 2020, some ads could have the shape of an actual \emph{Enriched result}. The URL also changed its color to gray and moved above the title. 

\begin{figure}[h]
    \centering
    \includegraphics[width=1.0\columnwidth]{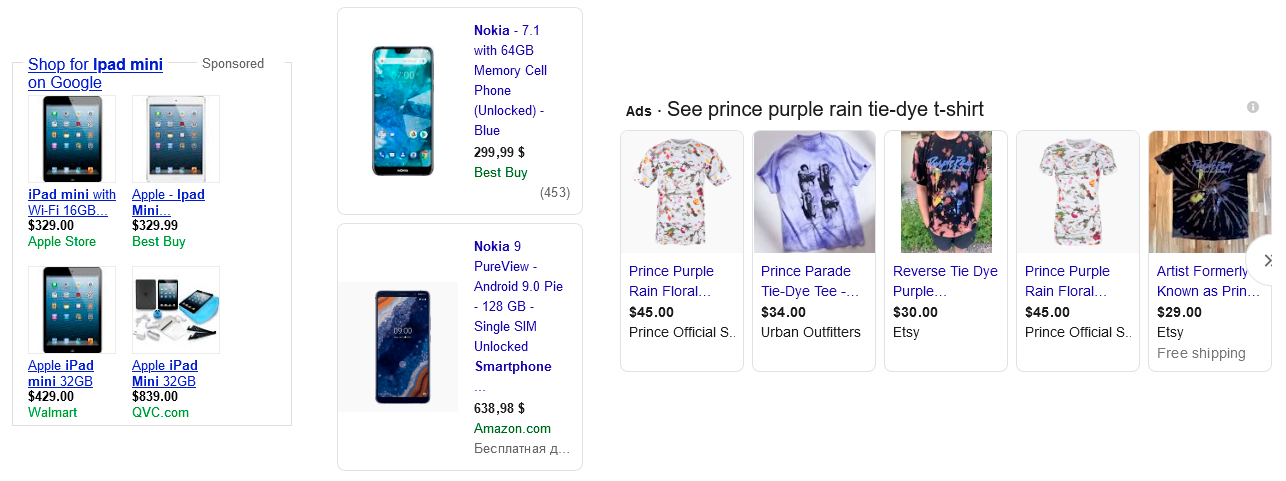}
    \caption{Shopping ads from 2013 (left), 2019 (center) and 2020 (right)}
    \label{fig:shoppingres}
\end{figure}

The other type of ads correspond to shopping content and are, therefore, called \textbf{shopping ads}, seen in Figure~\ref{fig:shoppingres}. These are activated when the query is commercial~\cite{Rosu2020}. These are very striking results, listing various information about each product. This element appeared for the first time in 2013, lasting until now. Its positioning is usually at the top of the results container and, more frequently, at the right of that container. Shopping ads used to be exclusive to the right sidebar, displaying one to four results in a matrix. In 2018, each result was embedded in an individual card. Until 2019, query terms were highlighted in bold. In 2020, this element appeared in the results container in a carousel of cards with more width and less height. The \emph{Cards} design pattern has been applied to this element since 2019, and the \emph{Carousel} since 2020. Cards display content composed of distinct parts, generally about a single subject, to form one coherent piece of content designed to expose information efficiently~\cite{Tidwell2020,10.1145/3291992.3292002}. It is usual for cards to accompany other cards, carrying similar content but addressing different subjects. The Carousel is a horizontal strip of simple cards, letting the user scroll horizontally to view them and encouraging the inspection of the following items~\cite{Tidwell2020,toxboe}. Since the element's beginning, the \emph{Thumbnail Grid} has been applied, enabling a quick overview of images by shrinking the original ones~\cite{Tidwell2020,10.1145/964696.964707}. 

In Figure~\ref{fig:patterns_all}, it is possible to identify two major advertisement areas in SERP: top ads and right ads, following a typical SERP layout~\cite[p.~489]{Baeza-yates2011}. 
Elements in the right sidebar move horizontally over time because some interfaces force this sidebar to be responsive to the screen's width. Top ads are the most common positions for ads within a soft vertical variation. Older interfaces placed advertisements on a fully colored div that occupied almost the entire viewport width for the same reason described in Section~\ref{sec:orgres}. Compared with regular results and other features, we can see in Figure~\ref{fig:patterns_all} that sponsored results are more concentrated in the visible area, probably motivated by the fact that searchers rarely scroll~\cite{Wilson2011}.

\subsection{Features} \label{sec:features_ind}
SERP features complement organic and sponsored results, attempting to provide answers to the query without just pointing to websites that might deliver that information.

Figure~\ref{fig:patterns_all}, in green, shows features spread around the interface, mainly in the visible area and upper half of the scrolling area. 
Many features are similar to regular results but with more significant height. These features also share a place in the sidebar with advertisements, recently becoming more present than the latter. Most frames with large height values correspond to the well-known \emph{Knowledge Panel}. A horizontal area is noticeable, generally assigned to the \emph{Carousel} and the \emph{Carousel Grid}. 

Due to space limitations, we selected a subset of features based on their lifetime and distinctiveness to be analyzed in detail. The \emph{Video Pack} feature will not be considered for its similarity with the \emph{Image Pack}. The same happens with \emph{Direct Answer Results}, \emph{Local Pack}, \emph{People Also Ask}, \emph{Carousel}, \emph{Carousel Grid}, \emph{Recipe Cards}, \emph{Twitter Pack}, \emph{Category Hierarchy}, and \emph{Covid-19 Left Panel} for their shorter presence. More information about these features is available in another paper~\cite{serpevolutionShort} and online\footnote{\url{https://bedgarone.github.io/serpevolution/elements}}.  

\begin{figure}[h]
    \centering
    \includegraphics[width=1.0\columnwidth]{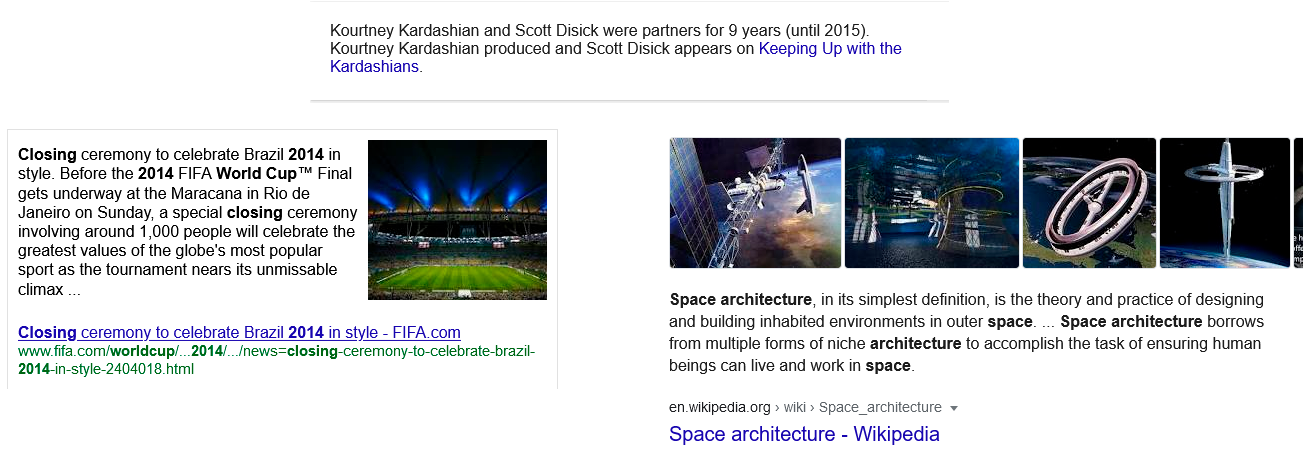}
    \caption{Featured Snippets from 2016 (top), 2017 (left) and 2020 (right)}
    \label{fig:fsnippets}
\end{figure}

\textbf{Featured Snippets}, seen in Figure \ref{fig:fsnippets} are answer boxes in which Google responds to a question-related query based on information taken from a page \cite{Vaughn2019}. This element appeared for the first time in 2016, lasting until now. Its positioning is consistently at the top of the results container. Although these snippets resemble enriched results, they differ in the type and position of extra content. In featured snippets, the additional content appears before the result's title and not after, as happens in enriched results. The content of a featured snippet was initially made of a short paragraph with answering information. It evolved to a general layout, used until now, consisting of a larger paragraph, a thumbnail at the upper-right corner, and the title and link to the information's source, to where it is possible to navigate. The answer also started to be returned as an ordered list or table. Instead of a thumbnail, a video or a carousel of images may also accompany it. In 2018, when available, Google introduced the date of the source's publication after the information paragraph. The underlined style of the title was removed in 2017, and the green URL was changed to a gray breadcrumb URL in 2020, as in the organic and sponsored results. Text styling is used in both paragraph and title, enhancing relevant words in bold. Studies have found that these snippets help make more accurate decisions when containing the correct information~\cite{Bink2022}.

\begin{figure}[h]
    \centering
    \includegraphics[width=0.7\columnwidth]{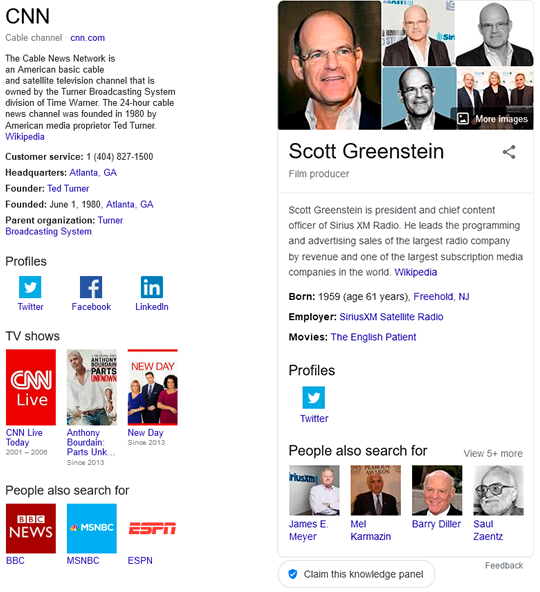}
    \caption{Knowledge Panel from 2016 (left) and 2018 (right)}
    \label{fig:kpanels}
\end{figure}

The \textbf{Knowledge Panel}, seen in Figure \ref{fig:kpanels}, is perhaps the highlight of SERP features. It is a dynamic feature that provides direct information in various formats within the same panel, pointing to related content. The contents range from text to images, ratings, social profiles, factual information, and similar search topics~\cite{Rosu2020}, helping the user to understand a particular subject quickly and facilitating a more in-depth search~\cite{DannySullivan2020}. This element appeared for the first time in 2014, lasting until now. Its positioning is always at the right of the results container. The basic structure consists of a panel with a top thumbnail of the subject, vertically followed by a title, a website link if applicable, a resume paragraph usually by \emph{Wikipedia}, a structured list of direct information, and a block of \emph{People also search for}. During the following years, Google introduced other content highly dependable on the search topic and variable in coverage and quality~\cite{Lurie2018}. It is common to have this panel populated by Wikipedia information~\cite{Vincent2021}. The graphics of this element was stable over time, following the improvements in Google's interface design. The \emph{Grid of Equals} and \emph{Thumbnail Grid} design patterns have been applied to this element since its beginning. Grid of Equals is a pattern to display items in a grid or matrix, each following a standard template, linking to respective pages~\cite{toxboe}. The \emph{Cards}, \emph{Carousel}, and \emph{Module Tabs} design patterns were applied to this element in 2018. 

\begin{figure}[h]
    \centering
    \includegraphics[width=1.0\columnwidth]{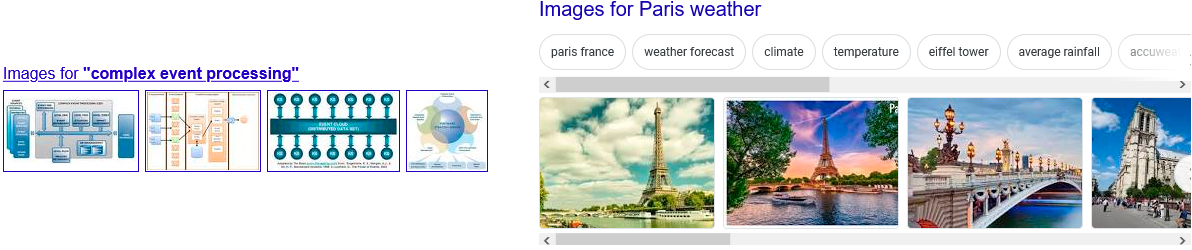}
    \caption{Image Pack from 2010 (left) and 2020 (right)}
    \label{fig:imagepack}
\end{figure}

The \textbf{Image Pack}, seen in Figure \ref{fig:imagepack}, presents a set of images taken from various sources in Google's index in searches where visual content is valuable~\cite{Vaughn2019}. This element appeared for the first time in 2006, but only after 2010 did it frequently appear, lasting until now. Its positioning is highly variable throughout the results container. Most of the time, the content was a title associating images with the search query and a block of image thumbnails. In 2014, Google included a link for `more images' and a link to report pictures. In 2019, Google introduced a bar of image categories. It could have simple buttons or buttons with a thumbnail associated with its category. The graphics started with a considerable presence of blue colors, typical in Google's early interfaces when images had a blue border. In 2014, Google removed this border, but the main change was in 2019 when images started to be in a carousel. In 2018, the layout of a matrix appeared for the first time. These changes increased the element's area in the last two years. In 2020 the title, as usual in most elements, turned dark gray. The carousel of image categories changed its shape to a line that expands to a matrix in the form of progressive disclosure using the \emph{Collapsible Panels} pattern. The \emph{Grid of Equals} design pattern has been applied to this element since 2018, while \emph{Thumbnail Grid}, naturally, since its beginning. Arguello et al.~\cite{Arguello2012} examined vertical results, such as images, in aggregated search. They found that these results had more clicks in more complex tasks and that users were divided in their preferences for vertical search displays.

\begin{figure}[h]
    \centering
    \includegraphics[width=1.0\columnwidth]{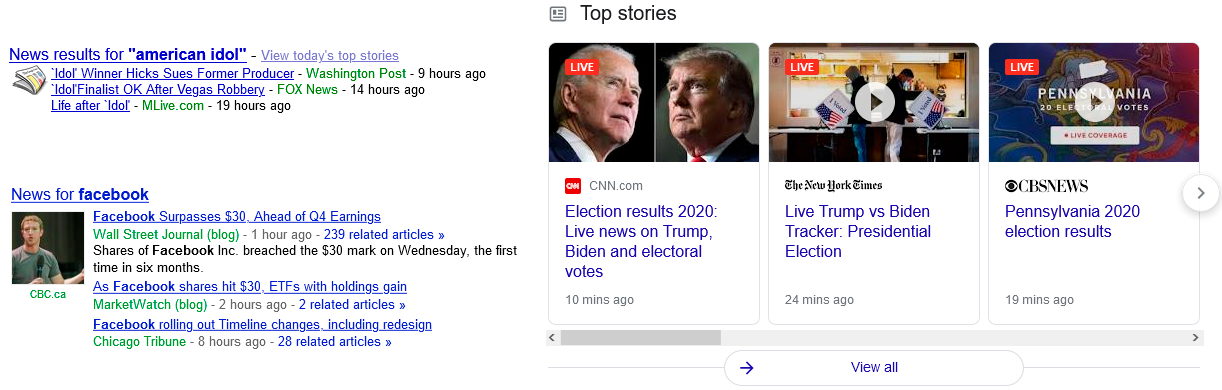}
    \caption{Top Stories from 2006 (left-top), 2013 (left-bottom) and 2020 (right)}
    \label{fig:newsres}
\end{figure}

The \textbf{Top Stories}, seen in Figure~\ref{fig:newsres}, are blocks of three or more recent news considered relevant to the query, recently placed in the form of a carousel~\cite{Rosu2020}. Each story is now presented with a thumbnail, publisher, and timestamp. This element appeared for the first time in 2004, but only after 2011 it started to appear frequently, lasting until now. Its positioning is mainly in the visible area of the results container. The element's content started with a vertical list of at most four news titles, each followed by the source's name and how long ago it was published. In 2006, a journal icon was placed at the left of the list, and a link to `today's top stories' was introduced. In 2013, the icon was substituted by a thumbnail for the first news result, being the most important news in the element. The latter was complemented with an extract of the news, while the rest stayed the same. In 2020, leading to a considerable increase in the element's area, the graphics was majorly altered to display the results in a carousel of cards. However, the content was simplified to only present, for each result, a thumbnail, title, source, and how long ago it was published. As usual, the color scheme was mainly blueish and filled with blue borders and underlines. These were removed after 2020 with the softer gray colors for additional information but still blue titles. The \emph{Streams and Feeds} design pattern has been applied to this element since its beginning. It defines a pattern to list time-sensitive items chronologically, combining the sources in one place~\cite{toxboe}. However, in this element, the relation to the query appears to be more relevant than publication time since it is no longer possible to observe any chronological order. The \emph{Cards} and \emph{Carousel} design patterns have been applied since 2020.

\begin{figure}[h]
    \centering
    \includegraphics[width=1.0\columnwidth]{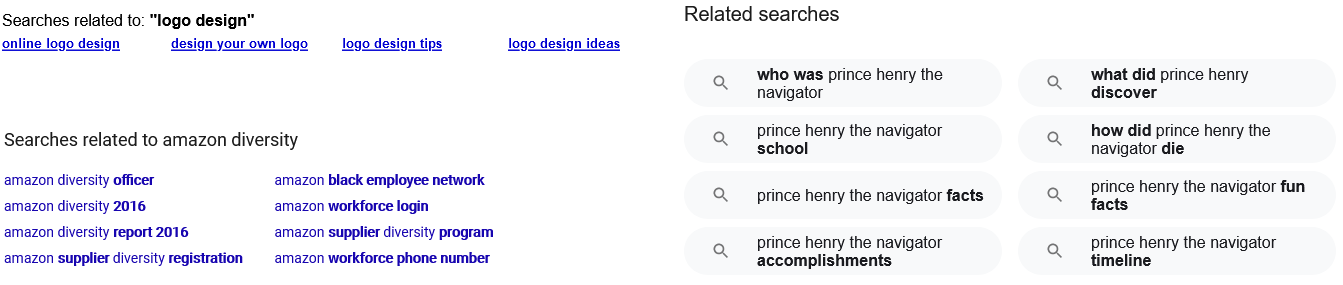}
    \caption{Related Searches from 2008 (left - top), 2017 (left - bottom) and 2020 (right)}
    \label{fig:relsearches}
\end{figure}

\textbf{Related Searches}, seen in Figure \ref{fig:relsearches}, is a common element on SERP pages from a very early age and offers suggestions for related searches, i.e., queries that are in some way related to the current query and may be good candidates for follow-on queries. These suggestions can be helpful to support exploration or provide query statements that express information needs in different ways~\cite{white2016}. Usually, these suggestions are generated based on search log data, either picking queries that frequently follow the current query~\cite{Jones2006} or clustering queries based on results' clicking~\cite{Craswell2007}. Each link takes the user to the respective SERP. This element appeared for the first time in 2008, lasting until now, except for 2010. Its positioning is always at the bottom of the results container. The content was diversified regarding how many suggestions would appear and its layout. Each suggestion of search is a hyperlinked title pointing to its respective SERP. Initially, it was organized in a matrix of columns. In 2011, Google reduced this schema to two columns, which could be displayed in just one column for suggestions with longer text. Until mid-2020, suggestions were blue and, until 2014, underlined. Google applied a search icon to each entry in 2020. Later, a new version changed the graphics, making each entry a button with a solid gray background, a search icon, and a title in black. This latter change contributed to a recent increase in the element's average area. 

\section{Aggregated Analysis}\label{sec:overallevol}

This section analyzes SERP's evolution from an aggregated perspective. 

\subsection{Elements' lifetime}

As shown, SERP have always had a large variety of elements, each with its evolution and active periods, as described in Section~\ref{sec:componentsevol}. According to the framework proposed by Wilson~\cite{Wilson2011}, \emph{input} features were analyzed in Section~\ref{sec:navinputs}, with the search box being the main one. To support \emph{control}, besides the input features, we also identified the \emph{related searches} feature. All the other analyzed features are \emph{informational}. It is important to note that, given the nature of this study, it was not possible to analyze dynamic features such as interactive querying or \emph{personalizable} features.


Black cells in Table~\ref{tab:presence} identify the years we detected elements in our dataset. For all years in-between black cells, we have conducted a manual search in other sources to avoid false negatives. In these cases, we have searched specific websites\footnote{\url{https://searchengineland.com}, \url{https://googlesystem.blogspot.com} and Internet Archive} for evidence of elements in such years. If found, we paint the cell grey. 
It is noticeable how SERP features have emerged in the last decade, contributing to a matrix full of element possibilities in recent years. Almost every SERP element includes well-known design patterns. A visual timeline with screenshots of these elements' presence in SERP is available online\footnote{\url{https://bedgarone.github.io/serpevolution/timeline/2010}}. 
 
\begin{table*}[h]
\footnotesize
\setlength\tabcolsep{2pt}
\caption{Presence of SERP elements from 2000 to 2020. In black the years in which the element appeared in our dataset. In grey are the years in which the element's existence is documented elsewhere.}
\begin{tabular}{@{}llllllllllllllllllllll@{}}
\toprule
& 2000 & 2001 & 2002 & 2003 & 2004 & 2005 & 2006 & 2007 & 2008 & 2009 & 2010 & 2011 & 2012 & 2013 & 2014 & 2015 & 2016 & 2017 & 2018 & 2019 & 2020\\ \midrule

\multicolumn{1}{l}{Visual Identity}     & \multicolumn{1}{l}{\cellcolor[HTML]{000000}} \ & \multicolumn{1}{l}{\cellcolor[HTML]{000000}} \ & \multicolumn{1}{l}{\cellcolor[HTML]{000000}} \ & \multicolumn{1}{l}{\cellcolor[HTML]{000000}} \ & \multicolumn{1}{l}{\cellcolor[HTML]{000000}} \ & \multicolumn{1}{l}{\cellcolor[HTML]{000000}} \ & \multicolumn{1}{l}{\cellcolor[HTML]{000000}} \ & \multicolumn{1}{l}{\cellcolor[HTML]{000000}} \ & \multicolumn{1}{l}{\cellcolor[HTML]{000000}} \ & \multicolumn{1}{l}{\cellcolor[HTML]{000000}} \ & \multicolumn{1}{l}{\cellcolor[HTML]{000000}} \ & \multicolumn{1}{l}{\cellcolor[HTML]{000000}} \ & \multicolumn{1}{l}{\cellcolor[HTML]{000000}} \ & \multicolumn{1}{l}{\cellcolor[HTML]{000000}} \ & \multicolumn{1}{l}{\cellcolor[HTML]{000000}} \ & \multicolumn{1}{l}{\cellcolor[HTML]{000000}} \ & \multicolumn{1}{l}{\cellcolor[HTML]{000000}} \ & \multicolumn{1}{l}{\cellcolor[HTML]{000000}} \ & \multicolumn{1}{l}{\cellcolor[HTML]{000000}} \ & \multicolumn{1}{l}{\cellcolor[HTML]{000000}} \ & \multicolumn{1}{l}{\cellcolor[HTML]{000000}} \\
\cmidrule(){2-22}

\multicolumn{1}{l}{Search Statistics}     & \multicolumn{1}{l}{\cellcolor[HTML]{000000}} \ & \multicolumn{1}{l}{\cellcolor[HTML]{000000}} \ & \multicolumn{1}{l}{\cellcolor[HTML]{000000}} \ & \multicolumn{1}{l}{\cellcolor[HTML]{000000}} \ & \multicolumn{1}{l}{\cellcolor[HTML]{000000}} \ & \multicolumn{1}{l}{\cellcolor[HTML]{000000}} \ & \multicolumn{1}{l}{\cellcolor[HTML]{000000}} \ & \multicolumn{1}{l}{\cellcolor[HTML]{000000}} \ & \multicolumn{1}{l}{\cellcolor[HTML]{000000}} \ & \multicolumn{1}{l}{\cellcolor[HTML]{000000}} \ & \multicolumn{1}{l}{\cellcolor[HTML]{000000}} \ & \multicolumn{1}{l}{\cellcolor[HTML]{000000}} \ & \multicolumn{1}{l}{\cellcolor[HTML]{000000}} \ & \multicolumn{1}{l}{\cellcolor[HTML]{000000}} \ & \multicolumn{1}{l}{\cellcolor[HTML]{000000}} \ & \multicolumn{1}{l}{\cellcolor[HTML]{000000}} \ & \multicolumn{1}{l}{\cellcolor[HTML]{000000}} \ & \multicolumn{1}{l}{\cellcolor[HTML]{000000}} \ & \multicolumn{1}{l}{\cellcolor[HTML]{000000}} \ & \multicolumn{1}{l}{\cellcolor[HTML]{000000}} \ & \multicolumn{1}{l}{\cellcolor[HTML]{000000}} \\
\cmidrule(){2-22}

\multicolumn{1}{l}{Navigation}     & \multicolumn{1}{l}{\cellcolor[HTML]{000000}} \ & \multicolumn{1}{l}{\cellcolor[HTML]{000000}} \ & \multicolumn{1}{l}{\cellcolor[HTML]{000000}} \ & \multicolumn{1}{l}{\cellcolor[HTML]{000000}} \ & \multicolumn{1}{l}{\cellcolor[HTML]{000000}} \ & \multicolumn{1}{l}{\cellcolor[HTML]{000000}} \ & \multicolumn{1}{l}{\cellcolor[HTML]{000000}} \ & \multicolumn{1}{l}{\cellcolor[HTML]{000000}} \ & \multicolumn{1}{l}{\cellcolor[HTML]{000000}} \ & \multicolumn{1}{l}{\cellcolor[HTML]{000000}} \ & \multicolumn{1}{l}{\cellcolor[HTML]{000000}} \ & \multicolumn{1}{l}{\cellcolor[HTML]{000000}} \ & \multicolumn{1}{l}{\cellcolor[HTML]{000000}} \ & \multicolumn{1}{l}{\cellcolor[HTML]{000000}} \ & \multicolumn{1}{l}{\cellcolor[HTML]{000000}} \ & \multicolumn{1}{l}{\cellcolor[HTML]{000000}} \ & \multicolumn{1}{l}{\cellcolor[HTML]{000000}} \ & \multicolumn{1}{l}{\cellcolor[HTML]{000000}} \ & \multicolumn{1}{l}{\cellcolor[HTML]{000000}} \ & \multicolumn{1}{l}{\cellcolor[HTML]{000000}} \ & \multicolumn{1}{l}{\cellcolor[HTML]{000000}} \\
\cmidrule(){2-22}

\multicolumn{1}{l}{User Inputs}     & \multicolumn{1}{l}{\cellcolor[HTML]{000000}} \ & \multicolumn{1}{l}{\cellcolor[HTML]{000000}} \ & \multicolumn{1}{l}{\cellcolor[HTML]{000000}} \ & \multicolumn{1}{l}{\cellcolor[HTML]{000000}} \ & \multicolumn{1}{l}{\cellcolor[HTML]{000000}} \ & \multicolumn{1}{l}{\cellcolor[HTML]{000000}} \ & \multicolumn{1}{l}{\cellcolor[HTML]{000000}} \ & \multicolumn{1}{l}{\cellcolor[HTML]{000000}} \ & \multicolumn{1}{l}{\cellcolor[HTML]{000000}} \ & \multicolumn{1}{l}{\cellcolor[HTML]{000000}} \ & \multicolumn{1}{l}{\cellcolor[HTML]{000000}} \ & \multicolumn{1}{l}{\cellcolor[HTML]{000000}} \ & \multicolumn{1}{l}{\cellcolor[HTML]{000000}} \ & \multicolumn{1}{l}{\cellcolor[HTML]{000000}} \ & \multicolumn{1}{l}{\cellcolor[HTML]{000000}} \ & \multicolumn{1}{l}{\cellcolor[HTML]{000000}} \ & \multicolumn{1}{l}{\cellcolor[HTML]{000000}} \ & \multicolumn{1}{l}{\cellcolor[HTML]{000000}} \ & \multicolumn{1}{l}{\cellcolor[HTML]{000000}} \ & \multicolumn{1}{l}{\cellcolor[HTML]{000000}} \ & \multicolumn{1}{l}{\cellcolor[HTML]{000000}} \\

\cmidrule(){2-22} 

\multicolumn{1}{l}{Regular Results}     & \multicolumn{1}{l}{\cellcolor[HTML]{000000}} \ & \multicolumn{1}{l}{\cellcolor[HTML]{000000}} \ & \multicolumn{1}{l}{\cellcolor[HTML]{000000}} \ & \multicolumn{1}{l}{\cellcolor[HTML]{000000}} \ & \multicolumn{1}{l}{\cellcolor[HTML]{000000}} \ & \multicolumn{1}{l}{\cellcolor[HTML]{000000}} \ & \multicolumn{1}{l}{\cellcolor[HTML]{000000}} \ & \multicolumn{1}{l}{\cellcolor[HTML]{000000}} \ & \multicolumn{1}{l}{\cellcolor[HTML]{000000}} \ & \multicolumn{1}{l}{\cellcolor[HTML]{000000}} \ & \multicolumn{1}{l}{\cellcolor[HTML]{000000}} \ & \multicolumn{1}{l}{\cellcolor[HTML]{000000}} \ & \multicolumn{1}{l}{\cellcolor[HTML]{000000}} \ & \multicolumn{1}{l}{\cellcolor[HTML]{000000}} \ & \multicolumn{1}{l}{\cellcolor[HTML]{000000}} \ & \multicolumn{1}{l}{\cellcolor[HTML]{000000}} \ & \multicolumn{1}{l}{\cellcolor[HTML]{000000}} \ & \multicolumn{1}{l}{\cellcolor[HTML]{000000}} \ & \multicolumn{1}{l}{\cellcolor[HTML]{000000}} \ & \multicolumn{1}{l}{\cellcolor[HTML]{000000}} \ & \multicolumn{1}{l}{\cellcolor[HTML]{000000}} \\
\cmidrule(){2-22} 

\multicolumn{1}{l}{Enriched Results} &
\multicolumn{1}{l}{} \ &
\multicolumn{1}{l}{} \ & 
\multicolumn{1}{l}{} \ &
\multicolumn{1}{l}{} \ &
\multicolumn{1}{l}{} \ & \multicolumn{1}{l}{} \ & \multicolumn{1}{l}{}        \ & \multicolumn{1}{l}{} \ & \multicolumn{1}{l}{\cellcolor[HTML]{000000}} \ & \multicolumn{1}{l}{\cellcolor[HTML]{000000}} \ & \multicolumn{1}{l}{\cellcolor[HTML]{e2e2e2}} \ & \multicolumn{1}{l}{\cellcolor[HTML]{e2e2e2}} \ & \multicolumn{1}{l}{\cellcolor[HTML]{000000}} \ & \multicolumn{1}{l}{\cellcolor[HTML]{000000}} \ & \multicolumn{1}{l}{\cellcolor[HTML]{000000}} \ & \multicolumn{1}{l}{\cellcolor[HTML]{000000}} \ & \multicolumn{1}{l}{\cellcolor[HTML]{000000}} \ & \multicolumn{1}{l}{\cellcolor[HTML]{000000}} \ & \multicolumn{1}{l}{\cellcolor[HTML]{000000}} \ & \multicolumn{1}{l}{\cellcolor[HTML]{000000}} \ & \multicolumn{1}{l}{\cellcolor[HTML]{000000}} \\

\cmidrule(){2-22} 

\multicolumn{1}{l}{Textual Ads}         & \multicolumn{1}{l}{\cellcolor[HTML]{000000}} \ & \multicolumn{1}{l}{}                         \ & \multicolumn{1}{l}{\cellcolor[HTML]{000000}} \ & \multicolumn{1}{l}{\cellcolor[HTML]{000000}} \ & \multicolumn{1}{l}{\cellcolor[HTML]{000000}} \ & \multicolumn{1}{l}{\cellcolor[HTML]{e2e2e2}} \ & \multicolumn{1}{l}{\cellcolor[HTML]{e2e2e2}} \ & \multicolumn{1}{l}{\cellcolor[HTML]{e2e2e2}} \ & \multicolumn{1}{l}{\cellcolor[HTML]{000000}} \ & \multicolumn{1}{l}{\cellcolor[HTML]{e2e2e2}} \ & \multicolumn{1}{l}{\cellcolor[HTML]{000000}} \ & \multicolumn{1}{l}{\cellcolor[HTML]{e2e2e2}} \ & \multicolumn{1}{l}{\cellcolor[HTML]{000000}} \ & \multicolumn{1}{l}{\cellcolor[HTML]{000000}} \ & \multicolumn{1}{l}{\cellcolor[HTML]{000000}} \ & \multicolumn{1}{l}{\cellcolor[HTML]{000000}} \ & \multicolumn{1}{l}{\cellcolor[HTML]{000000}} \ & \multicolumn{1}{l}{\cellcolor[HTML]{000000}} \ & \multicolumn{1}{l}{\cellcolor[HTML]{000000}} \ & \multicolumn{1}{l}{\cellcolor[HTML]{000000}} \ & \multicolumn{1}{l}{\cellcolor[HTML]{000000}} \\

\cmidrule(){2-22} 

\multicolumn{1}{l}{Shopping Ads}        & \multicolumn{1}{l}{}                         \ & \multicolumn{1}{l}{}                         \ & \multicolumn{1}{l}{}                         \ & \multicolumn{1}{l}{}                         \ & \multicolumn{1}{l}{}                         \ & \multicolumn{1}{l}{}                         \ & \multicolumn{1}{l}{}                         \ & \multicolumn{1}{l}{}                         \ & \multicolumn{1}{l}{}                         \ & \multicolumn{1}{l}{}                         \ & \multicolumn{1}{l}{}                         \ & \multicolumn{1}{l}{}                         \ & \multicolumn{1}{l}{}                         \ & \multicolumn{1}{l}{\cellcolor[HTML]{000000}} \ & \multicolumn{1}{l}{\cellcolor[HTML]{e2e2e2}} \ & \multicolumn{1}{l}{\cellcolor[HTML]{000000}} \ & \multicolumn{1}{l}{\cellcolor[HTML]{000000}} \ & \multicolumn{1}{l}{\cellcolor[HTML]{000000}} \ & \multicolumn{1}{l}{\cellcolor[HTML]{000000}} \ & \multicolumn{1}{l}{\cellcolor[HTML]{000000}} \ & \multicolumn{1}{l}{\cellcolor[HTML]{000000}} \\ \cmidrule(){2-22}

\multicolumn{1}{l}{Knowledge Panel}  & \multicolumn{1}{l}{}                         \ & \multicolumn{1}{l}{}                         \ & \multicolumn{1}{l}{}                         \ & \multicolumn{1}{l}{}                         \ & \multicolumn{1}{l}{}                         \ & \multicolumn{1}{l}{}                         \ & \multicolumn{1}{l}{}                         \ & \multicolumn{1}{l}{}                         \ & \multicolumn{1}{l}{}                         \ & \multicolumn{1}{l}{}                         \ & \multicolumn{1}{l}{}                         \ & \multicolumn{1}{l}{}                         \ & \multicolumn{1}{l}{}                         \ & \multicolumn{1}{l}{}                         \ & \multicolumn{1}{l}{\cellcolor[HTML]{000000}} \ & \multicolumn{1}{l}{\cellcolor[HTML]{000000}} \ & \multicolumn{1}{l}{\cellcolor[HTML]{000000}} \ & \multicolumn{1}{l}{\cellcolor[HTML]{000000}} \ & \multicolumn{1}{l}{\cellcolor[HTML]{000000}} \ & \multicolumn{1}{l}{\cellcolor[HTML]{000000}} \ & \multicolumn{1}{l}{\cellcolor[HTML]{000000}} \\ \cmidrule(){2-22} 
\multicolumn{1}{l}{Featured Snippets}   & \multicolumn{1}{l}{}                         \ & \multicolumn{1}{l}{}                         \ & \multicolumn{1}{l}{}                         \ & \multicolumn{1}{l}{}                         \ & \multicolumn{1}{l}{}                         \ & \multicolumn{1}{l}{}                         \ & \multicolumn{1}{l}{}                         \ & \multicolumn{1}{l}{}                         \ & \multicolumn{1}{l}{}                         \ & \multicolumn{1}{l}{}                         \ & \multicolumn{1}{l}{}                         \ & \multicolumn{1}{l}{}                         \ & \multicolumn{1}{l}{}                         \ & \multicolumn{1}{l}{}                         \ & \multicolumn{1}{l}{}                         \ & \multicolumn{1}{l}{}                         \ & \multicolumn{1}{l}{\cellcolor[HTML]{000000}} \ & \multicolumn{1}{l}{\cellcolor[HTML]{000000}} \ & \multicolumn{1}{l}{\cellcolor[HTML]{000000}} \ & \multicolumn{1}{l}{\cellcolor[HTML]{000000}} \ & \multicolumn{1}{l}{\cellcolor[HTML]{000000}} \\ \cmidrule(){2-22} 
\multicolumn{1}{l}{Direct Answer}       & \multicolumn{1}{l}{}                         \ & \multicolumn{1}{l}{}                         \ & \multicolumn{1}{l}{}                         \ & \multicolumn{1}{l}{}                         \ & \multicolumn{1}{l}{}                         \ & \multicolumn{1}{l}{}                         \ & \multicolumn{1}{l}{}                         \ & \multicolumn{1}{l}{}                         \ & \multicolumn{1}{l}{}                         \ & \multicolumn{1}{l}{}                         \ & \multicolumn{1}{l}{}                         \ & \multicolumn{1}{l}{}                         \ & \multicolumn{1}{l}{}                         \ & \multicolumn{1}{l}{}                         \ & \multicolumn{1}{l}{}                         \ & \multicolumn{1}{l}{}                         \ & \multicolumn{1}{l}{}                         \ & \multicolumn{1}{l}{}                         \ & \multicolumn{1}{l}{\cellcolor[HTML]{000000}} \ & \multicolumn{1}{l}{\cellcolor[HTML]{000000}} \ & \multicolumn{1}{l}{\cellcolor[HTML]{000000}} \\ \cmidrule(){2-22} 
\multicolumn{1}{l}{Local Pack}          & \multicolumn{1}{l}{}                         \ & \multicolumn{1}{l}{}                         \ & \multicolumn{1}{l}{}                         \ & \multicolumn{1}{l}{}                         \ & \multicolumn{1}{l}{}                         \ & \multicolumn{1}{l}{}                         \ & \multicolumn{1}{l}{}                         \ & \multicolumn{1}{l}{}                         \ & \multicolumn{1}{l}{}                         \ & \multicolumn{1}{l}{\cellcolor[HTML]{000000}} \ & \multicolumn{1}{l}{\cellcolor[HTML]{e2e2e2}} \ & \multicolumn{1}{l}{\cellcolor[HTML]{e2e2e2}} \ & \multicolumn{1}{l}{\cellcolor[HTML]{e2e2e2}} \ & \multicolumn{1}{l}{\cellcolor[HTML]{000000}} \ & \multicolumn{1}{l}{\cellcolor[HTML]{e2e2e2}} \ & \multicolumn{1}{l}{\cellcolor[HTML]{e2e2e2}} \ & \multicolumn{1}{l}{\cellcolor[HTML]{000000}} \ & \multicolumn{1}{l}{\cellcolor[HTML]{e2e2e2}} \ & \multicolumn{1}{l}{\cellcolor[HTML]{000000}} \ & \multicolumn{1}{l}{\cellcolor[HTML]{000000}} \ & \multicolumn{1}{l}{\cellcolor[HTML]{000000}} \\ \cmidrule(){2-22} 
\multicolumn{1}{l}{Image Pack}          \ & \multicolumn{1}{l}{}                         \ & \multicolumn{1}{l}{}                         \ & \multicolumn{1}{l}{}                         \ & \multicolumn{1}{l}{}                         \ & \multicolumn{1}{l}{}                         \ & \multicolumn{1}{l}{}                         \ & \multicolumn{1}{l}{\cellcolor[HTML]{000000}} \ & \multicolumn{1}{l}{}                         \ & \multicolumn{1}{l}{}                         \ & \multicolumn{1}{l}{}                         \ & \multicolumn{1}{l}{\cellcolor[HTML]{000000}} \ & \multicolumn{1}{l}{\cellcolor[HTML]{000000}} \ & \multicolumn{1}{l}{\cellcolor[HTML]{e2e2e2}} \ & \multicolumn{1}{l}{\cellcolor[HTML]{e2e2e2}} \ & \multicolumn{1}{l}{\cellcolor[HTML]{000000}} \ & \multicolumn{1}{l}{\cellcolor[HTML]{000000}} \ & \multicolumn{1}{l}{\cellcolor[HTML]{000000}} \ & \multicolumn{1}{l}{\cellcolor[HTML]{000000}} \ & \multicolumn{1}{l}{\cellcolor[HTML]{000000}} \ & \multicolumn{1}{l}{\cellcolor[HTML]{000000}} \ & \multicolumn{1}{l}{\cellcolor[HTML]{000000}} \\ \cmidrule(){2-22} 
\multicolumn{1}{l}{Video Pack}          & \multicolumn{1}{l}{}                         \ & \multicolumn{1}{l}{}                         \ & \multicolumn{1}{l}{}                         \ & \multicolumn{1}{l}{}                         \ & \multicolumn{1}{l}{}                         \ & \multicolumn{1}{l}{}                         \ & \multicolumn{1}{l}{}                         \ & \multicolumn{1}{l}{}                         \ & \multicolumn{1}{l}{}                         \ & \multicolumn{1}{l}{}                         \ & \multicolumn{1}{l}{\cellcolor[HTML]{000000}} \ & \multicolumn{1}{l}{\cellcolor[HTML]{000000}} \ & \multicolumn{1}{l}{}                         \ & \multicolumn{1}{l}{}                         \ & \multicolumn{1}{l}{}                         \ & \multicolumn{1}{l}{\cellcolor[HTML]{000000}} \ & \multicolumn{1}{l}{\cellcolor[HTML]{000000}} \ & \multicolumn{1}{l}{\cellcolor[HTML]{000000}} \ & \multicolumn{1}{l}{\cellcolor[HTML]{000000}} \ & \multicolumn{1}{l}{\cellcolor[HTML]{000000}} \ & \multicolumn{1}{l}{\cellcolor[HTML]{000000}} \\ \cmidrule(){2-22} 
\multicolumn{1}{l}{Top Stories}         & \multicolumn{1}{l}{}                         \ & \multicolumn{1}{l}{}                         \ & \multicolumn{1}{l}{}                         \ & \multicolumn{1}{l}{}                         \ & \multicolumn{1}{l}{\cellcolor[HTML]{000000}} \ & \multicolumn{1}{l}{}                         \ & \multicolumn{1}{l}{\cellcolor[HTML]{000000}} \ & \multicolumn{1}{l}{\cellcolor[HTML]{e2e2e2}} \ & \multicolumn{1}{l}{}                         \ & \multicolumn{1}{l}{}                         \ & \multicolumn{1}{l}{}                         \ & \multicolumn{1}{l}{\cellcolor[HTML]{000000}} \ & \multicolumn{1}{l}{\cellcolor[HTML]{000000}} \ & \multicolumn{1}{l}{\cellcolor[HTML]{000000}} \ & \multicolumn{1}{l}{\cellcolor[HTML]{000000}} \ & \multicolumn{1}{l}{\cellcolor[HTML]{000000}} \ & \multicolumn{1}{l}{\cellcolor[HTML]{000000}} \ & \multicolumn{1}{l}{\cellcolor[HTML]{000000}} \ & \multicolumn{1}{l}{\cellcolor[HTML]{e2e2e2}} \ & \multicolumn{1}{l}{\cellcolor[HTML]{000000}} \ & \multicolumn{1}{l}{\cellcolor[HTML]{000000}} \\ \cmidrule(){2-22} 
\multicolumn{1}{l}{Carousel}            & \multicolumn{1}{l}{}                         \ & \multicolumn{1}{l}{}                         \ & \multicolumn{1}{l}{}                         \ & \multicolumn{1}{l}{}                         \ & \multicolumn{1}{l}{}                         \ & \multicolumn{1}{l}{}                         \ & \multicolumn{1}{l}{}                         \ & \multicolumn{1}{l}{}                         \ & \multicolumn{1}{l}{}                         \ & \multicolumn{1}{l}{}                         \ & \multicolumn{1}{l}{}                         \ & \multicolumn{1}{l}{}                         \ & \multicolumn{1}{l}{}                         \ & \multicolumn{1}{l}{}                         \ & \multicolumn{1}{l}{}                         \ & \multicolumn{1}{l}{\cellcolor[HTML]{000000}} \ & \multicolumn{1}{l}{\cellcolor[HTML]{000000}} \ & \multicolumn{1}{l}{\cellcolor[HTML]{000000}} \ & \multicolumn{1}{l}{\cellcolor[HTML]{000000}} \ & \multicolumn{1}{l}{}                         \ & \multicolumn{1}{l}{}                         \\ \cmidrule(){2-22} 
\multicolumn{1}{l}{Carousel Grid}       & \multicolumn{1}{l}{}                         \ & \multicolumn{1}{l}{}                         \ & \multicolumn{1}{l}{}                         \ & \multicolumn{1}{l}{}                         \ & \multicolumn{1}{l}{}                         \ & \multicolumn{1}{l}{}                         \ & \multicolumn{1}{l}{}                         \ & \multicolumn{1}{l}{}                         \ & \multicolumn{1}{l}{}                         \ & \multicolumn{1}{l}{}                         \ & \multicolumn{1}{l}{}                         \ & \multicolumn{1}{l}{}                         \ & \multicolumn{1}{l}{}                         \ & \multicolumn{1}{l}{}                         \ & \multicolumn{1}{l}{}                         \ & \multicolumn{1}{l}{}                         \ & \multicolumn{1}{l}{}                         \ & \multicolumn{1}{l}{\cellcolor[HTML]{000000}} \ & \multicolumn{1}{l}{\cellcolor[HTML]{000000}} \ & \multicolumn{1}{l}{}                         \ & \multicolumn{1}{l}{\cellcolor[HTML]{000000}} \\ \cmidrule(){2-22} 
\multicolumn{1}{l}{People Also Ask}     & \multicolumn{1}{l}{}                         \ & \multicolumn{1}{l}{}                         \ & \multicolumn{1}{l}{}                         \ & \multicolumn{1}{l}{}                         \ & \multicolumn{1}{l}{}                         \ & \multicolumn{1}{l}{}                         \ & \multicolumn{1}{l}{}                         \ & \multicolumn{1}{l}{}                         \ & \multicolumn{1}{l}{}                         \ & \multicolumn{1}{l}{}                         \ & \multicolumn{1}{l}{}                         \ & \multicolumn{1}{l}{}                         \ & \multicolumn{1}{l}{}                         \ & \multicolumn{1}{l}{}                         \ & \multicolumn{1}{l}{}                         \ & \multicolumn{1}{l}{}                         \ & \multicolumn{1}{l}{\cellcolor[HTML]{000000}} \ & \multicolumn{1}{l}{\cellcolor[HTML]{000000}} \ & \multicolumn{1}{l}{\cellcolor[HTML]{000000}} \ & \multicolumn{1}{l}{\cellcolor[HTML]{000000}} \ & \multicolumn{1}{l}{\cellcolor[HTML]{000000}} \\ \cmidrule(){2-22} 
\multicolumn{1}{l}{Related Searches}    & \multicolumn{1}{l}{}                         \ & \multicolumn{1}{l}{}                         \ & \multicolumn{1}{l}{}                         \ & \multicolumn{1}{l}{}                         \ & \multicolumn{1}{l}{}                         \ & \multicolumn{1}{l}{}                         \ & \multicolumn{1}{l}{}                         \ & \multicolumn{1}{l}{}                         \ & \multicolumn{1}{l}{\cellcolor[HTML]{000000}} \ & \multicolumn{1}{l}{\cellcolor[HTML]{000000}} \ & \multicolumn{1}{l}{\cellcolor[HTML]{e2e2e2}} \ & \multicolumn{1}{l}{\cellcolor[HTML]{000000}} \ & \multicolumn{1}{l}{\cellcolor[HTML]{000000}} \ & \multicolumn{1}{l}{\cellcolor[HTML]{000000}} \ & \multicolumn{1}{l}{\cellcolor[HTML]{000000}} \ & \multicolumn{1}{l}{\cellcolor[HTML]{000000}} \ & \multicolumn{1}{l}{\cellcolor[HTML]{000000}} \ & \multicolumn{1}{l}{\cellcolor[HTML]{000000}} \ & \multicolumn{1}{l}{\cellcolor[HTML]{000000}} \ & \multicolumn{1}{l}{\cellcolor[HTML]{000000}} \ & \multicolumn{1}{l}{\cellcolor[HTML]{000000}} \\ \cmidrule(){2-22} 
\multicolumn{1}{l}{Twitter Pack}        & \multicolumn{1}{l}{}                         \ & \multicolumn{1}{l}{}                         \ & \multicolumn{1}{l}{}                         \ & \multicolumn{1}{l}{}                         \ & \multicolumn{1}{l}{}                         \ & \multicolumn{1}{l}{}                         \ & \multicolumn{1}{l}{}                         \ & \multicolumn{1}{l}{}                         \ & \multicolumn{1}{l}{}                         \ & \multicolumn{1}{l}{}                         \ & \multicolumn{1}{l}{}                         \ & \multicolumn{1}{l}{}                         \ & \multicolumn{1}{l}{}                         \ & \multicolumn{1}{l}{}                         \ & \multicolumn{1}{l}{}                         \ & \multicolumn{1}{l}{\cellcolor[HTML]{000000}} \ & \multicolumn{1}{l}{\cellcolor[HTML]{000000}} \ & \multicolumn{1}{l}{\cellcolor[HTML]{000000}} \ & \multicolumn{1}{l}{\cellcolor[HTML]{000000}} \ & \multicolumn{1}{l}{\cellcolor[HTML]{000000}} \ & \multicolumn{1}{l}{\cellcolor[HTML]{000000}} \\ \cmidrule(){2-22} 
\multicolumn{1}{l}{Recipe Cards}        & \multicolumn{1}{l}{}                         \ & \multicolumn{1}{l}{}                         \ & \multicolumn{1}{l}{}                         \ & \multicolumn{1}{l}{}                         \ & \multicolumn{1}{l}{}                         \ & \multicolumn{1}{l}{}                         \ & \multicolumn{1}{l}{}                         \ & \multicolumn{1}{l}{}                         \ & \multicolumn{1}{l}{}                         \ & \multicolumn{1}{l}{}                         \ & \multicolumn{1}{l}{}                         \ & \multicolumn{1}{l}{}                         \ & \multicolumn{1}{l}{}                         \ & \multicolumn{1}{l}{}                         \ & \multicolumn{1}{l}{}                         \ & \multicolumn{1}{l}{}                         \ & \multicolumn{1}{l}{}                         \ & \multicolumn{1}{l}{}                         \ & \multicolumn{1}{l}{}                         \ & \multicolumn{1}{l}{}                         \ & \multicolumn{1}{l}{\cellcolor[HTML]{000000}} \\ \cmidrule(){2-22} 
\multicolumn{1}{l}{Category Hierarchy}  & \multicolumn{1}{l}{\cellcolor[HTML]{000000}}                         \ & \multicolumn{1}{l}{\cellcolor[HTML]{000000}}                         \ & \multicolumn{1}{l}{\cellcolor[HTML]{000000}}                         \ & \multicolumn{1}{l}{\cellcolor[HTML]{000000}} \ & \multicolumn{1}{l}{\cellcolor[HTML]{000000}} \ & \multicolumn{1}{l}{}                         \ & \multicolumn{1}{l}{}                         \ & \multicolumn{1}{l}{}                         \ & \multicolumn{1}{l}{}                         \ & \multicolumn{1}{l}{}                         \ & \multicolumn{1}{l}{}                         \ & \multicolumn{1}{l}{}                         \ & \multicolumn{1}{l}{}                         \ & \multicolumn{1}{l}{}                         \ & \multicolumn{1}{l}{}                         \ & \multicolumn{1}{l}{}                         \ & \multicolumn{1}{l}{}                         \ & \multicolumn{1}{l}{}                         \ & \multicolumn{1}{l}{}                         \ & \multicolumn{1}{l}{}                         \ & \multicolumn{1}{l}{}                         \\ \cmidrule(){2-22} 
\multicolumn{1}{l}{Covid-19 Left Panel} & \multicolumn{1}{l}{}                         \ & \multicolumn{1}{l}{}                         \ & \multicolumn{1}{l}{}                         \ & \multicolumn{1}{l}{}                         \ & \multicolumn{1}{l}{}                         \ & \multicolumn{1}{l}{}                         \ & \multicolumn{1}{l}{}                         \ & \multicolumn{1}{l}{}                         \ & \multicolumn{1}{l}{}                         \ & \multicolumn{1}{l}{}                         \ & \multicolumn{1}{l}{}                         \ & \multicolumn{1}{l}{}                         \ & \multicolumn{1}{l}{}                         \ & \multicolumn{1}{l}{}                         \ & \multicolumn{1}{l}{}                         \ & \multicolumn{1}{l}{}                         \ & \multicolumn{1}{l}{}                         \ & \multicolumn{1}{l}{}                         \ & \multicolumn{1}{l}{}                         \ & \multicolumn{1}{l}{}                         \ & \multicolumn{1}{l}{\cellcolor[HTML]{000000}} \\ \bottomrule
\end{tabular}
\label{tab:presence}
\end{table*}

\subsection{Design pattern application}

Table~\ref{tab:googpatterns} maps each element with the patterns proposed by Tidwell et al.~\cite{Tidwell2020}, along with the start date of that appliance. Older SERP elements make later use of design patterns for individual improvement. In contrast, some contemporary elements may have arisen from the need to apply a design pattern solution whose traces are evident from the element's beginning. The website\footnote{\url{https://bedgarone.github.io/serpevolution/patterns}} lists design patterns with images and elements that use them.


\newcolumntype{C}[1]{>{\centering\let\newline\\\arraybackslash\hspace{0pt}}m{#1}}
\begin{table*}[h]
\footnotesize
\setlength\tabcolsep{2pt}
\caption{Design Patterns and time of their appliance to SERP elements. Cells without a dash after the year represent single years.}
\begin{tabular}{@{}lC{1cm}C{1cm}C{1cm}C{1cm}C{1cm}C{1cm}C{1cm}C{1cm}C{1cm}C{1cm}C{1cm}@{}}
\toprule
\multicolumn{1}{l}{} & \multicolumn{1}{c}{Organizing} & \multicolumn{1}{c}{\cellcolor[HTML]{F0F0F0}{Navigation}} & \multicolumn{4}{c}{Layout} & \multicolumn{4}{c}{\cellcolor[HTML]{F0F0F0}{Lists}} & \multicolumn{1}{c}{Input} \\

\cmidrule(){2-12} 
 & Streams and Feeds \ & {\cellcolor[HTML]{F0F0F0}{Breadcrumbs}} \ & Grid of Equals \ & Module Tabs \ & Accordion \ & Collapsible Panels \ & {\cellcolor[HTML]{F0F0F0}{Cards}} \ & {\cellcolor[HTML]{F0F0F0}{Thumbnail Grid}} \ & {\cellcolor[HTML]{F0F0F0}{Carousel}} \ & {\cellcolor[HTML]{F0F0F0}{Pagination}} \ & Input Prompt \\ \midrule

Visual Identity  & & & & & & & & & & & \\\cmidrule(){2-12} 
Search Statistics  & & & & & & & & & & & \\\cmidrule(){2-12} 

Navigation  & & & & & & & & & &\multicolumn{1}{c}{\cellcolor[HTML]{000000}{\color[HTML]{FFFFFF} 2000-2020}} & \\ \cmidrule(){2-12} 
User Inputs & & & & & & & & & & & \multicolumn{1}{c}{\cellcolor[HTML]{000000}{\color[HTML]{FFFFFF} 2000-2020}} \\ \cmidrule(){2-12} 

Regular Results  & & \multicolumn{1}{c}{\cellcolor[HTML]{000000}{\color[HTML]{FFFFFF} 2019-}} & & & \multicolumn{1}{c}{{\color[HTML]{FFFFFF} }} & & & & & & \\ \cmidrule(){2-12} 
Enriched Results  & & \multicolumn{1}{c}{\cellcolor[HTML]{000000}{\color[HTML]{FFFFFF} 2020-}} & & & \multicolumn{1}{c}{{\color[HTML]{FFFFFF} 2020-}} & & & & & & \\\cmidrule(){2-12} 
Textual Ads  & & & & & & & & & & & \\ \cmidrule(){2-12} 
Shopping Ads  & & & & & & & \multicolumn{1}{c}{\cellcolor[HTML]{000000}{\color[HTML]{FFFFFF} 2019-}}\ & \multicolumn{1}{c}{}\ & \multicolumn{1}{c}{\cellcolor[HTML]{000000}{\color[HTML]{FFFFFF} 2020-}} & & \\ \cmidrule(){2-12} 
 
Knowledge Panel  & & & \multicolumn{1}{c}{\cellcolor[HTML]{000000}{\color[HTML]{FFFFFF} 2014-}} \ & \multicolumn{1}{c}{\cellcolor[HTML]{000000}{\color[HTML]{FFFFFF} 2018}} & & & \multicolumn{1}{c}{\cellcolor[HTML]{000000}{\color[HTML]{FFFFFF} 2018}}\ & \multicolumn{1}{c}{\cellcolor[HTML]{000000}{\color[HTML]{FFFFFF} 2014-}}\ & \multicolumn{1}{c}{\cellcolor[HTML]{000000}{\color[HTML]{FFFFFF} 2016-}} & & \\\cmidrule(){2-12} 
Featured Snippets  & & & & & & & & \multicolumn{1}{c}{\cellcolor[HTML]{000000}{\color[HTML]{FFFFFF} 2016-}}\ & \multicolumn{1}{c}{\cellcolor[HTML]{000000}{\color[HTML]{FFFFFF} 2020-}} & & \\\cmidrule(){2-12} 
Direct Answer  & & \multicolumn{1}{c}{\cellcolor[HTML]{000000}{\color[HTML]{FFFFFF} 2018-}} & & & \multicolumn{1}{c}{{\color[HTML]{FFFFFF} 2018-}} & & & \multicolumn{1}{c}{} & & & \\\cmidrule(){2-12} 
Local Pack  & & & & & & & & & & & \\\cmidrule(){2-12} 

Image Pack  & & & \multicolumn{1}{c}{\cellcolor[HTML]{000000}{\color[HTML]{FFFFFF} 2018-}} & & & \multicolumn{1}{c}{\cellcolor[HTML]{000000}{\color[HTML]{FFFFFF} 2019-}} & & \multicolumn{1}{c}{\cellcolor[HTML]{000000}{\color[HTML]{FFFFFF} 2006-}}\ & \multicolumn{1}{c}{\cellcolor[HTML]{000000}{\color[HTML]{FFFFFF} 2019-}} & & \\\cmidrule(){2-12} 
Video Pack  & & & & & & & \multicolumn{1}{c}{\cellcolor[HTML]{000000}{\color[HTML]{FFFFFF} 2015-}} & \multicolumn{1}{c}{} & \multicolumn{1}{c}{\cellcolor[HTML]{000000}{\color[HTML]{FFFFFF} 2015-}} & & \\\cmidrule(){2-12} 
Top Stories  & \multicolumn{1}{c}{\cellcolor[HTML]{000000}{\color[HTML]{FFFFFF} 2004-}} & & & & & & \multicolumn{1}{c}{\cellcolor[HTML]{000000}{\color[HTML]{FFFFFF} 2020-}} & \multicolumn{1}{c}{} & \multicolumn{1}{c}{\cellcolor[HTML]{000000}{\color[HTML]{FFFFFF} 2020-}} & & \multicolumn{1}{c}{} \\\cmidrule(){2-12} 
Carousel  & & \multicolumn{1}{c}{\cellcolor[HTML]{000000}{\color[HTML]{FFFFFF} 2015-}} \ & & & & & \multicolumn{1}{c}{\cellcolor[HTML]{000000}{\color[HTML]{FFFFFF} 2016-}} \ & \multicolumn{1}{c}{\cellcolor[HTML]{000000}{\color[HTML]{FFFFFF} 2015-}}\ & \multicolumn{1}{c}{\cellcolor[HTML]{000000}{\color[HTML]{FFFFFF} 2015-}} & & \\\cmidrule(){2-12} 
Carousel Grid  & & \multicolumn{1}{c}{\cellcolor[HTML]{000000}{\color[HTML]{FFFFFF} 2017-}} \ & \multicolumn{1}{c}{\cellcolor[HTML]{000000}{\color[HTML]{FFFFFF} 2017-}} & & & & \multicolumn{1}{c}{\cellcolor[HTML]{000000}{\color[HTML]{FFFFFF} 2017-}} & \multicolumn{1}{c}{} & & & \\\cmidrule(){2-12} 

People Also Ask  & & & & & \multicolumn{1}{c}{\cellcolor[HTML]{000000}{\color[HTML]{FFFFFF} 2016-}} & & & & & & \\\cmidrule(){2-12} 
Related Searches  & & & & & & & & & & & \\\cmidrule(){2-12} 
Twitter Pack  & & & & & & & \multicolumn{1}{c}{\cellcolor[HTML]{000000}{\color[HTML]{FFFFFF} 2017-}} & \multicolumn{1}{c}{} & \multicolumn{1}{c}{\cellcolor[HTML]{000000}{\color[HTML]{FFFFFF} 2017-}} & & \\\cmidrule(){2-12} 
Recipe Cards  & & & & & & \multicolumn{1}{c}{\cellcolor[HTML]{000000}{\color[HTML]{FFFFFF} 2020-}}\ & \multicolumn{1}{c}{\cellcolor[HTML]{000000}{\color[HTML]{FFFFFF} 2020-}} & \multicolumn{1}{c}{} & & & \\\cmidrule(){2-12} 
Category Hierarchy  & & \multicolumn{1}{c}{\cellcolor[HTML]{000000}{\color[HTML]{FFFFFF} 2000-2004}} & & & & & & & & \multicolumn{1}{c}{} & \\\cmidrule(){2-12} 
Covid-19 Left Panel  & & & & & & & & & & \multicolumn{1}{c}{} & \\

\bottomrule
\end{tabular}
\label{tab:googpatterns}
\end{table*}

\subsection{Highlights}

\begin{figure*}[h]
    \centering
    \includegraphics[width=0.85\textwidth]{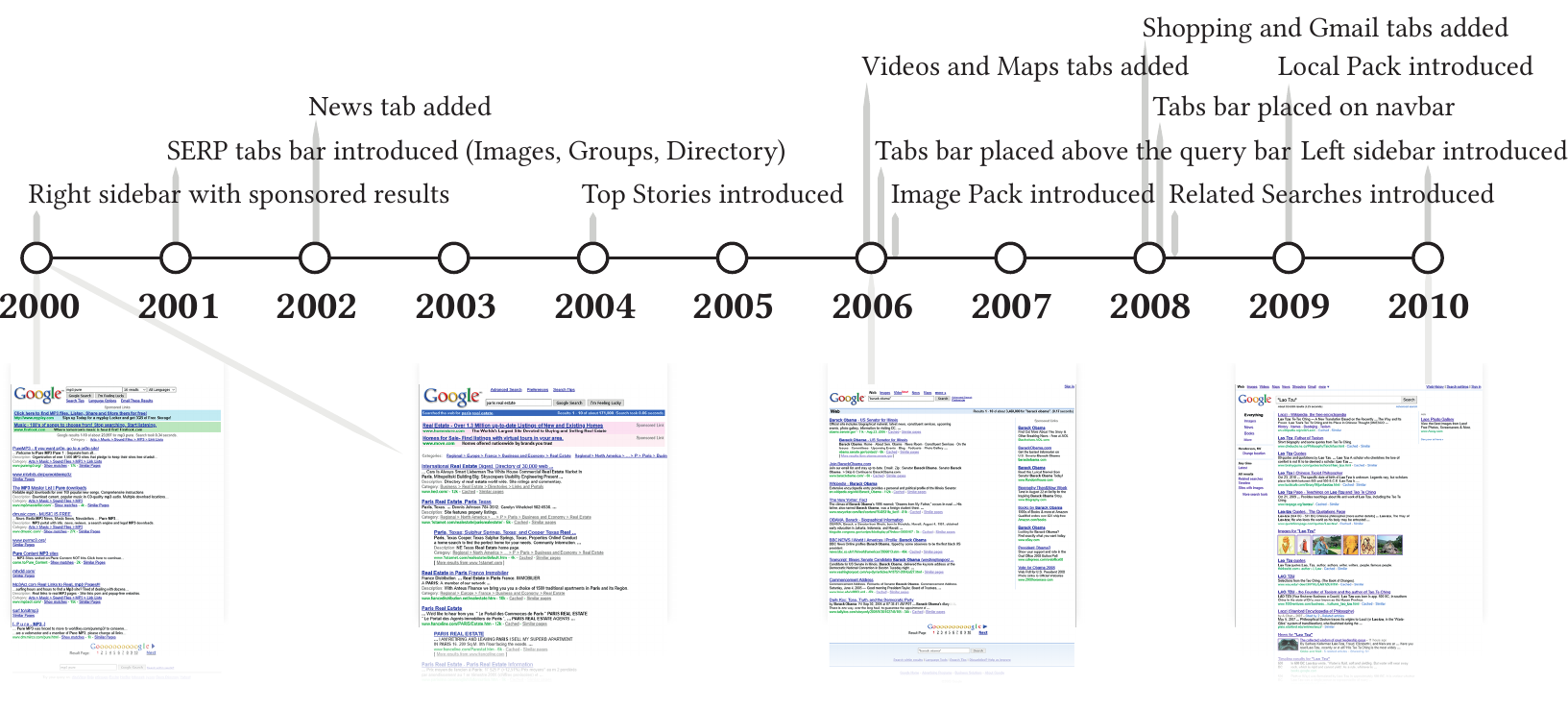}
    \caption{Highlights of SERP overall evolution (top) and Interfaces' visual evolution (bottom) from 2000 to 2010}
    \label{fig:globaltimeline1}
\end{figure*}

\begin{figure*}[h]
    \centering
    \includegraphics[width=0.85\textwidth]{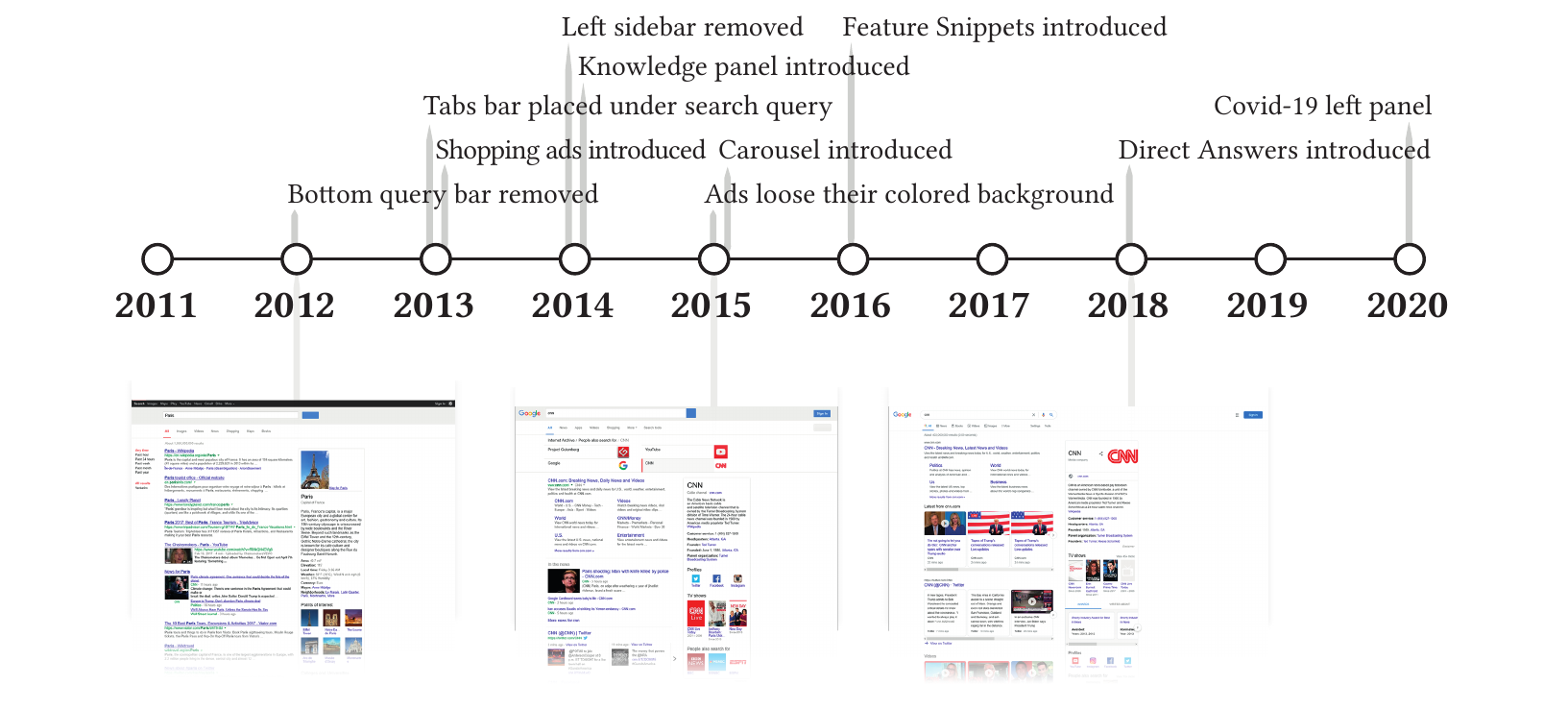}
    \caption{Highlights of SERP overall evolution (top) and Interfaces' visual evolution (bottom) from 2011 to 2020}
    \label{fig:globaltimeline2}
\end{figure*}

We identify the most relevant changes along the upper part of a two-decade timeline of SERP in Figures~\ref{fig:globaltimeline1} and \ref{fig:globaltimeline2}. These changes correspond to the entry of new elements or significant changes in input and navigation options. This timeline of changes is also available online\footnote{\url{https://bedgarone.github.io/serpevolution/timeline}}, enhanced with images and the option to filter the entries for navigation changes or element additions. 

As stated before, the first stage of the analysis used a visual selection from a monthly sample from the dataset. Figures~\ref{fig:globaltimeline1} and \ref{fig:globaltimeline2} also display, in their lower part, how Google's SERP interfaces evolved in terms of design. In these timelines, we include the main versions of the SERP interface based on significant changes. Apart from how the overall interfaces visually evolved, similar timelines about visual identity, search statistics, and navigation are available on the website\footnote{\url{https://bedgarone.github.io/serpevolution/design}}.

Google's initial interfaces had few depth levels in their HTML DOM. The first interface design, traced in 2000, differentiated the sponsored results with a colored background. A second search query bar existed at the bottom of the page, and the user could change the number of results presented. Google removed this bar in the following interfaces. The one traced from 2000 to 2004 revealed a right block of results, exclusive to sponsored ones. It marked the appearance of the first bar with tabs directing the user to other types of content (e.g., images and news). The fourth interface design, traced from 2010 to 2012, was not left-oriented but varied in a spaced manner depending on the screen's width. It introduced a sidebar on the left, containing tabs to manage the results, but some of these tabs were duplicated due to the navbar's tabs bar mentioned in Section~\ref{sec:navinputs}.

Significant aesthetic changes occurred in 2012. The fifth interface design relates to the launch of the Knowledge Graph, with the right column being divided between it and sponsored results. Some modifications were found earlier in the dataset during those years, the sixth interface design. However, a design close to the current one began at the end of 2018, the seventh interface design. As noted in some elements' graphics, this interface focused on modernizing its lines.

\subsection{User interface area}

\begin{figure}[h]
    \centering
    \includegraphics[width=0.7\columnwidth]{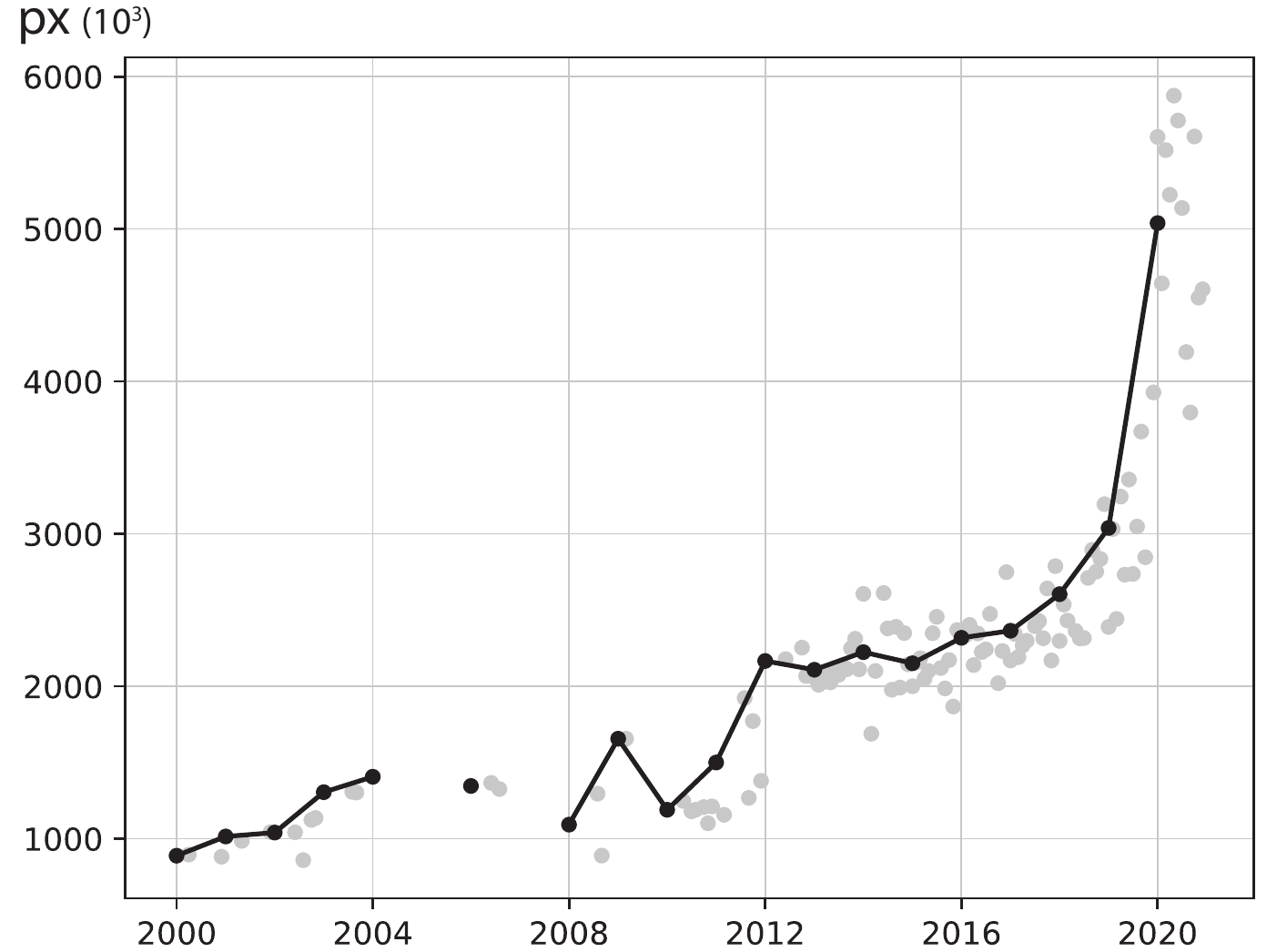}
    \caption{Interface area evolution, measured in squared pixel units}
    \label{fig:interfacearea}
\end{figure}

We calculated the area of all screenshots in the dataset to analyze its evolution over time. Figure~\ref{fig:interfacearea} shows the development of interface area per month (dots) and per year (line), measured in pixels. Each entry in the chart corresponds to the average area per month for all captures in the dataset. Months without values are months without captures in the dataset, as indicated in Table~\ref{tab:dataset}. Results show an increase close to exponential due to the appearance of SERP features that have added extra content to SERP, thus, making them more extensive over time. We can also notice this evolution in the animated overlaying of elements on the website\footnote{Available at \url{https://bedgarone.github.io/serpevolution/layout}}.  

\subsection{Files' size and number}

\begin{figure}[h]
    \centering
    \includegraphics[width=1.0\columnwidth]{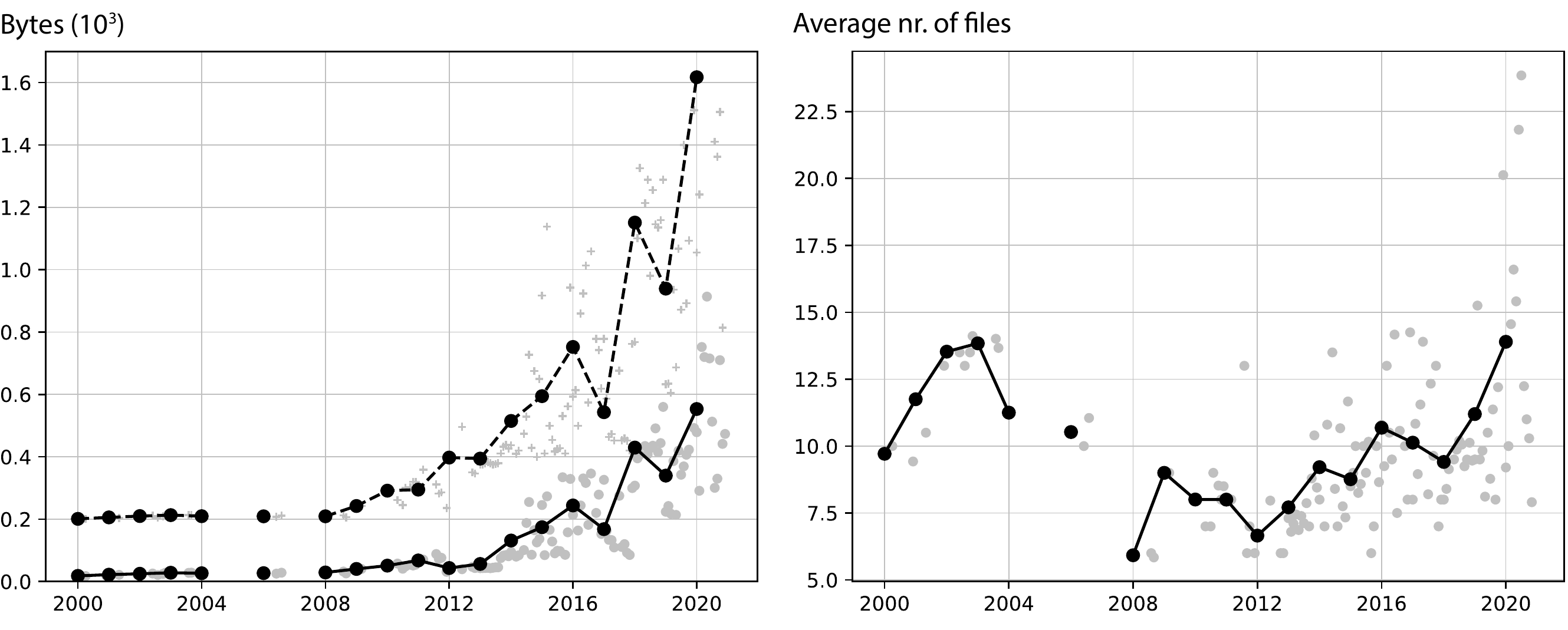}
    \caption{Source code size (left - line - circumference), Source code + Files folder size (left - dashed - plus sign) and average number of files associated (right) over time}
    \label{fig:filezsize}
\end{figure}

We made a similar approach to study the variation in file size regarding the entire dataset. Figure~\ref{fig:filezsize} represents the SERP captures' file size evolution and the number of files in its folder. For each capture, we summed the size of each associated file and averaged it per month and, consequently, year. We did not consider embedded files in the HTML but files linked through the associated files folder.

Size results accompany the development of the interface area, as seen in Figure~\ref{fig:interfacearea}, expressing a steep rise in the last few years. This increase cannot be related to a surge in the associated files, as later values share similar values with the first years of SERP existence. Nevertheless, results suggest that SERP sought to reduce the number of related files, achieving this aim during the first decade. This number started to rise again because interfaces evolved and demanded more images and graphics, which can increase the files needed to load a SERP. Besides, protocol advancements such as HTTP pipelining or SPDY may have contributed to the increase of these associated assets.

\section{Discussion} \label{sec:discussion}

SERP are no longer pages with ``10 blue links''. We have shown that the range of SERP elements has been growing continuously but steeper in the last few years. Although this increase might contribute to cluttering the page and falling into the ``more is less'' trap, we expect SERP to continue evolving quickly~\cite[p.~512]{Baeza-yates2011}. 
Interfaces have kept track of web development's evolution, as shown by the regular adoption of design patterns. 

The \emph{Category Hierarchy} stands out as the only feature used in the first years and later discontinued. The decrease in web directories' popularity may have been the reason for this. 

Although organic results have been keeping a relatively stable format, SERP have become more diversified over time, providing increasingly sophisticated navigational aids to enhance results interpretation, relevance assessment, and user satisfaction~\cite{Chakrabarti2009, Teevan2009, Haas2011}. These features complement organic and sponsored results and attempt to infer users' needs and quickly satisfy them even if not explicitly mentioned in their search queries~\cite{Chakrabarti2009} following a ``universal search'' vision~\cite{UniversalSearch2007}. As shown, most SERP elements are informational, providing information about results. 

The growth in SERP elements almost exponentially increased interface area since modern pages need more vertical space. 
These pages are getting heavier. Surprisingly, given the number of current features that use non-textual content, the average number of files is not much more significant than in 2000. Notwithstanding, we noticed a higher dispersion in the average number of files in more recent years, as seen in Figure~\ref{fig:filezsize}.

Aggregating results from heterogeneous sources - verticals - and presenting them in a single interface – aggregated search – has become standard practice~\cite{Zhou2012}. We could notice this in the \emph{Image Pack}, \emph{Video Pack}, \emph{Local Pack}, and \emph{Top Stories} features. Other features like the \emph{Featured Snippet} assemble information in different formats extracted from one source. 
Since 2020, information on the Web has dramatically increased in quantity and diversity. Videos are nowadays much more popular than they were at the beginning of the century, and content from new platforms such as location technology (e.g., Google Maps) or social media (e.g., Twitter) has emerged. This evolution naturally affected the need for the change of the SERP elements. The fact that graphical information is processed before the textual information~\cite{Hochstotter2009} might also explain the recurrent appearance of images and videos in features. 
On the other hand, the evolution of SERP is strongly informed and influenced by users' search behavior. Users scarcely look at results other than the first ones~\cite{10.1145/1076034.1076063} and tend to reformulate the query if they cannot find promising results at the top of the list~\cite{Guan2007}. They rarely look at results, such as videos or news, in their respective SERP `tabs'~\cite{Sullivan2003}. This behavior may also motivate search engines to include aggregations of other types of results~\cite{Hochstotter2009}, not for the diversity, because most users will not reach them outside the first results page.

The SERP are also affected by the interests of the search engine providers who provide users not only with relevant results but also with results of their interest. This reality gained prominence when the European Commission concluded that Google abused its market dominance by the way it presented sponsored results~\cite{EUPressRelease}. Although not a focus of analysis of this work, it is frequent to see Google showing results from its maps service, YouTube results in the video container, shopping results from its shopping ads service, and blurring the lines between organic and sponsored results. These decisions have a higher impact on users with less search engine knowledge, who are more likely to trust and use Google~\cite{LewaMisplaced}.

SERP features often allow the user to interact with the contents of a web page directly from the SERP~\cite{10.5555/1824082,Haas2011}. This cannibalizes clicks~\cite{Chilton2011} and might mean that users get satisfied without clicking on search results, which was defined by Li et al.~\cite{Li2009} as ``good abandonment''. 
Studies have found that features that provide direct answers improve user engagement on SERP, reduce user effort, and promote user satisfaction~\cite{Wu2020}. Besides contributing to user satisfaction, these features also encourage user engagement and, thus, revenue~\cite{10.1145/2487575.2488217, Haas2011}.

Our results reinforce the idea that evaluation measures solely based on the list of ``10 blue links'' must be rethought based on the SERP we have today. The standard practices of aggregating results from heterogeneous verticals and including features that provide direct answers on SERP have implications for how users interact with search systems and, therefore, on their evaluation. The cannibalization of clicks requests evaluations that consider other types of interactions with the SERP. Challenges emerge in the way users' feedback is explored, either explicitly from user studies or implicitly from weblogs. Work has already been conducted to rethink evaluation in the context of aggregated search pages~\cite{Zhou2012} and good abandonment scenarios~\cite{Khabsa2016}.

\section{Conclusions and Future Work} \label{sec:conclusion}
Using Google as a case study, we studied how SERP user interfaces evolved over two decades. While existing research has relied on the actual states of these interfaces, we have updated and improved the analysis with an evolution perspective, addressing old and new elements, their positioning, size, and patterns. We extracted and provide a dataset with 5,000+ SERP captures, including HTML versions and screenshots. 

We showed that SERP are becoming more diverse in terms of elements, aggregating content from different verticals and including more features that provide direct answers. These changes affect user behavior that, more often, abandon the page satisfied, the so-called ``good abandonment''.

In the future, we want to analyze other web search engines' SERP and compare results. We would also like to explore the evolution of SERP in mobile environments. Here, we would like to know if the increase in the SERP area found in this work results in a more significant differentiation of user interfaces between desktop and mobile environments. As stated previously, features that require interaction with the SERP were not analyzed here because our page captures from the Internet Archive don't allow such interaction. Given the importance of such features, we would like to explore them in current SERP versions. Studies that analyze SERP characteristics by types of queries (e.g., informational, navigational) and user studies comparing old and contemporary SERP would also be interesting.

\begin{acks}
The Master in Informatics and Computing Engineering and the Department of Informatics Engineering of the Faculty of Engineering of the University of Porto supported this work by funding the registration fee.
\end{acks}
 
\bibliographystyle{ACM-Reference-Format}
\bibliography{diss-bibliography}

\end{document}